\ifx\pdfoutput\undefined\else\pdfoutput=1\fi
\documentclass[ASNA,twocolumn]{USG} 
\usepackage{anyfontsize} %
\usepackage{colortbl}   
\usepackage{newtxtext}
\usepackage{graphicx}
\usepackage{textcomp}
\usepackage{xcolor}
\usepackage{url}
\usepackage{tabularx}
\usepackage{adjustbox}
\usepackage{nameref}
\usepackage{longtable}
\usepackage{multirow}
\usepackage{threeparttable}
\usepackage{supertabular,booktabs}
\usepackage{cuted}
\usepackage{caption}
\usepackage{amsmath}
\usepackage{subcaption}
\usepackage{pgfplots}
\pgfplotsset{compat=1.18}
\usetikzlibrary{patterns,arrows.meta,decorations.pathreplacing,calc}
\usepackage{footnote}
\usepackage{siunitx} 
\usepackage{steinmetz}
\DeclareCaptionLabelSeparator{wileybar}{\space|\space}
\graphicspath{{./images/}}
\specialissue{Preprint}
\articletype{SPECIAL ISSUE PAPER (PREPRINT)}%
\subarticletype{Submitted manuscript}
\received{}
\revised{}
\journal{Concurrency and Computation: Practice and Experience}
\volume{}
\copyyear{2026}
\startpage{1}
\articledoi{}

\ifx\pdfminorversion\undefined\else\pdfminorversion=7\fi
\historydates{}
\begin{document}
\title{Design and Evaluation of Energy-Efficient Whisper Dot-Product Kernel Offloading on a CGLA Architecture}
\transtitle{Design and Evaluation of Energy-Efficient Whisper Dot-Product Kernel Offloading on a CGLA Architecture}
\author[1]{Takuto Ando}[https://orcid.org/0009-0005-1873-869X]
\author[1]{Yu Eto}[https://orcid.org/0009-0006-0640-1683]
\author[1]{Ayumu Takeuchi}[https://orcid.org/0009-0006-2094-3163]
\author[1]{Yasuhiko Nakashima}[https://orcid.org/0000-0002-9457-5061]

\authormark{ANDO \textsc{et al.}}
\titlemark{WHISPER DOT-PRODUCT KERNEL OFFLOADING ON A CGLA ARCHITECTURE}

\address{\orgdiv{Graduate School of Science and Technology, }\orgname{Nara Institute of Science and Technology, }%
\orgaddress{\state{Nara}, \country{Japan}.}} 

\corres{Takuto Ando  (\email{antaku7585@gmail.com})}



\fundingInfo{This work was supported by the JST-ALCA-Next Program (Grant Number JPMJAN23F4) and JSPS KAKENHI (Grant No. 22H00515). We also acknowledge the activities of VDEC, The University of Tokyo, in collaboration with NIHON SYNOPSYS G.K.}


\transkeywords{Automatic Speech Recognition | Whisper | hardware/software co-design | CGLA | IMAX}

\newcommand{\varPDP}{\text{PDP}}
\newcommand{\varEDP}{\text{EDP}}
\newcommand{\varT}[1]{\text{T}_{\text{#1}}}
\newcommand{\varP}[1]{\text{P}_{\text{#1}}}

\captionsetup[figure]{labelsep=wileybar}
\captionsetup[table]{labelsep=wileybar}

\makeatletter
\def\fnum@figure{{\bf\MakeUppercase{F\kern1.5\@p@t I\kern1.5\@p@t G\kern1.5\@p@t U\kern1.5\@p@t R\kern1.5\@p@t E\kern1.5\@p@t}}~\textbf{\thefigure}}
\def\fnum@table{{\bf\MakeUppercase{T\kern1.5\@p@t A\kern1.5\@p@t B\kern1.5\@p@t L\kern1.8\@p@t E\kern1.5\@p@t}}~\thetable}
\makeatother

\abstract[ABSTRACT]{%
    In this paper, we implement and evaluate Whisper dot-product kernel offloading on IMAX, a programmable Coarse-Grained Linear Arrays~(CGLAs) architecture.
    Whisper-tiny.en profiling on an ARM Cortex-A72 shows that dot-product operations account for \SI{90.6}{\percent} of FP16 execution time and \SI{87.1}{\percent} of Q8\_0 execution time.
    To address this kernel bottleneck, we combine kernel mapping, local-memory sizing, and burst scheduling.
    The implementation uses inline FP16-to-FP32 conversion, 2-way SIMD FMA on a 64-bit datapath, column-wise multithreading, and mixed execution in which aligned vector segments run on IMAX and residual segments run concurrently on the host CPU.
    We evaluate the design with an FPGA prototype and a \SI{28}{\nm} ASIC projection at \SI{840}{\mega\hertz}.
    For Whisper-tiny.en, 32\,KB local memory and burst length~16 jointly minimize PDP and EDP.
    Under a TDP-based cross-platform comparison, the projected IMAX records a PDP of \SI{11.58}{\joule} for Whisper-tiny.en Q8\_0, 2.35$\times$ lower than Jetson AGX Orin (\SI{27.16}{\joule}) and 10.48$\times$ lower than RTX~4090 (\SI{121.38}{\joule}).
    The same design extends to Whisper-base.en and Whisper-small.en, where the PDP gap narrows as 32\,KB local-memory coverage drops from \SI{93.8}{\percent} for tiny to about \SI{66.5}{\percent} for base and small.
    These results position IMAX as a programmable architecture for lower-PDP local ASR in the tiny-model regime.}

 \transabstract[transABSTRACT]{In this paper, we implement and evaluate Whisper dot-product kernel offloading on IMAX, a programmable Coarse-Grained Linear Arrays~(CGLAs) architecture. Whisper-tiny.en profiling on an ARM Cortex-A72 shows that dot-product operations account for \SI{90.6}{\percent} of FP16 execution time and \SI{87.1}{\percent} of Q8\_0 execution time. To address this kernel bottleneck, we combine kernel mapping, local-memory sizing, and burst scheduling. The implementation uses inline FP16-to-FP32 conversion, 2-way SIMD FMA on a 64-bit datapath, column-wise multithreading, and mixed execution in which aligned vector segments run on IMAX and residual segments run concurrently on the host CPU. We evaluate the design with an FPGA prototype and a \SI{28}{\nm} ASIC projection at \SI{840}{\mega\hertz}. For Whisper-tiny.en, 32\,KB local memory and burst length~16 jointly minimize PDP and EDP. Under a TDP-based cross-platform comparison, the projected IMAX records a PDP of \SI{11.58}{\joule} for Whisper-tiny.en Q8\_0, 2.35$\times$ lower than Jetson AGX Orin (\SI{27.16}{\joule}) and 10.48$\times$ lower than RTX~4090 (\SI{121.38}{\joule}). The same design extends to Whisper-base.en and Whisper-small.en, where the PDP gap narrows as 32\,KB local-memory coverage drops from \SI{93.8}{\percent} for tiny to about \SI{66.5}{\percent} for base and small. These results position IMAX as a programmable architecture for lower-PDP local ASR in the tiny-model regime.}

\abbr{AI, Artificial Intelligence; ASR, Automatic Speech Recognition; ASIC, Application-Specific Integrated Circuit; CER, Character Error Rate; CGLA, Coarse-Grained Linear Array; CGRA, Coarse-Grained Reconfigurable Array; DMA, Direct Memory Access; FPGA, Field-Programmable Gate Array; FMA, Fused Multiply-Add; GPU, Graphics Processing Unit; GPGPU, General-Purpose Graphics Processing Unit; LLM, Large Language Model; LMM, Local Memory Module; MAC, Multiply-Accumulate; PDP, Power-Delay Product; PE, Processing Element; SIMD, Single Instruction Multiple Data; SoC, System-on-Chip; SOTA, State-of-the-Art; TDP, Thermal Design Power; WER, Word Error Rate.}





\maketitle


\section{Introduction}

Large Language Models~(LLMs)\nobreak\cite{expl_llm,healai,llm_mas,touvron2023llamaopenefficientfoundation} and Automatic Speech Recognition~(ASR) systems\nobreak\cite{ASR_1} are deployed across production services worldwide.
Whisper\nobreak\cite{radford2022robustspeechrecognitionlargescale} is an open encoder-decoder ASR model used in smart assistants\nobreak\cite{ASR_smart_voice_assistant}, real-time transcription\nobreak\cite{ASR_T2T}, and medical documentation\nobreak\cite{ASR_emergency_medicine}.

These workloads depend heavily on General-Purpose Graphics Processing Units~(GPGPUs), and the resulting energy cost is growing.
The International Energy Agency~(IEA) projects that data center electricity consumption could double by 2030, reaching approximately 945~TWh\nobreak\cite{iea_energy_ai,shehabi2024united}.
That level slightly exceeds Japan's total annual electricity consumption.
Scaling AI inference on a single general-purpose architecture cannot keep pace with this growth in power budget.

Specialized hardware, such as Application-Specific Integrated Circuits~(ASICs) and Field-Programmable Gate Arrays~(FPGAs), addresses this gap by trading generality for lower energy per operation.
Coarse-Grained Reconfigurable Arrays~(CGRAs) occupy an intermediate point. They map dataflow graphs directly onto an array of processing elements~(PEs), reducing memory-bound stalls while retaining programmability.
The Coarse-Grained Linear Arrays~(CGLAs) architecture, IMAX\nobreak\cite{imax_access}, extends this idea by targeting the trade-off between ASIC-level power and GPGPU-level programmability.
IMAX linearly arranges PEs and Local Memory Modules~(LMMs), absorbing irregular memory access patterns through single-cycle local reads and hardware-managed double-buffering.
In this paper, we study Whisper as a practical open ASR workload for architecture evaluation.
Whisper serves not as a proxy for every commercial ASR product, but as an open encoder-decoder Transformer that permits controlled comparison across hardware platforms under identical local-execution conditions.
Proprietary smartphone TPU deployments and cloud ASR services mix closed models, software stacks, network effects, and system-level optimizations, all of which would obscure the hardware-level effects we seek to isolate.

\vspace{0.5em}
\begin{strip}
\centering
\includegraphics[width=0.96\textwidth]{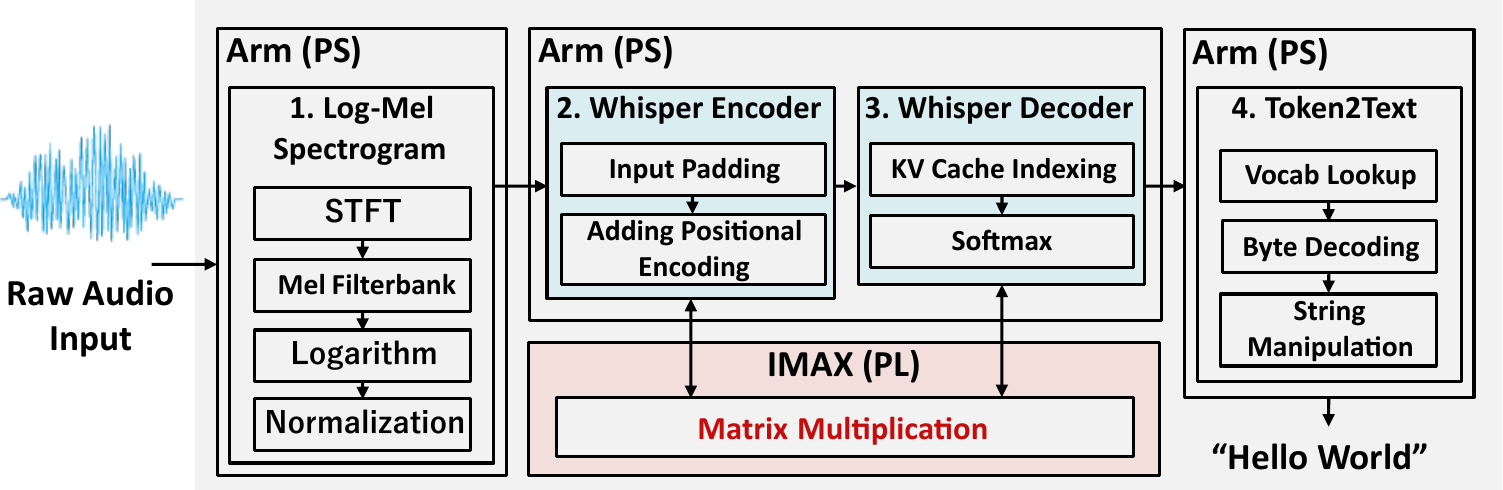}
\captionof{figure}{The Whisper ASR processing flow. Raw audio is first converted to an 80-channel log-mel spectrogram, which the encoder processes through a convolutional front-end and stacked Transformer layers to produce context embeddings. The decoder then autoregressively generates output tokens by attending to both the encoder output and previously generated tokens.}
\label{fig:whisper_process}
\end{strip}
\vspace{0.25em}
We implement and evaluate the primary computational kernel of Whisper on IMAX, a CGLA accelerator targeting power-constrained edge devices.
Local inference on embedded hardware eliminates data transmission and cloud dependency\nobreak\cite{HAN2026100674}, and is required in privacy-sensitive and connectivity-limited settings such as medical documentation, edge terminals, and offline transcription.
Fig.~\ref{fig:whisper_process} shows the ASR processing flow using the Whisper model.
The encoder and decoder consume most of the computation, with the dot-product kernel alone dominating CPU execution time.
We offload this kernel to IMAX and study two architectural parameters, LMM size and burst length.
The primary target is Whisper-tiny.en, from which the proposed operating points are derived.
The same analysis extends to Whisper-base.en and Whisper-small.en to expose where the tiny-oriented design scales and where architectural bottlenecks emerge.
Full acceleration of every Whisper operator is out of scope. The evaluation focuses on how dot-product offloading affects end-to-end behavior under a fixed whisper.cpp software stack.

To quantify the bottleneck motivating this work, we profiled Whisper-tiny.en running on the host CPU.
The dot-product operation accounts for \SI{90.6}{\percent} and \SI{87.1}{\percent} of total CPU execution time for the FP16 and Q8\_0 models, respectively, while all remaining operations, including Softmax and Layer Normalization, collectively account for less than \SI{13}{\percent}.
According to Amdahl's Law, offloading this single operation alone yields a theoretical maximum system-level speedup of \num{10.6}$\times$ for FP16 and \num{7.8}$\times$ for Q8\_0, assuming perfect acceleration of the targeted kernel.
These numbers motivate the IMAX-based dot-product accelerator and set the upper bound against which measured system-level speedup is compared.

This paper extends our previous conference version on Whisper ASR acceleration.
The conference paper established the basic feasibility of Whisper acceleration on IMAX for the tiny model.
This journal version extends that study in four ways:
\begin{itemize}
\item \textbf{CPU Profiling of Whisper-tiny.en.} We add CPU profiling and an Amdahl's-Law-based interpretation of the dominant kernel.
\item \textbf{Burst-Length and LMM-Size Sensitivity.} We add sensitivity studies for both burst length and LMM size.
\item \textbf{Extension to Base and Small Models.} We extend the evaluation from Whisper-tiny.en to Whisper-base.en and Whisper-small.en to characterize where the tiny-model design continues to reduce PDP and where it does not.
\item \textbf{Discussion of LMM Coverage and Power.} We add a broader discussion of limitations and the LMM power-capacity trade-off that appears beyond the tiny regime.
\end{itemize}

The main contributions of this paper are as follows:
\begin{itemize}
\item \textbf{Whisper Dot-Product Mapping on IMAX.} IMAX offloads the dominant Whisper dot-product workload under controlled local-execution conditions by mapping the dynamic and variable-length dot products of the encoder-decoder model onto a programmable datapath.
\item \textbf{32\,KB LMM and Burst Length~16 for Whisper-tiny.en.} 32\,KB LMM and burst length~16 jointly minimize PDP and EDP for Whisper-tiny.en under the \SI{28}{\nm} power model. We then apply the same design to Whisper-base.en and Whisper-small.en.
\item \textbf{LMM Coverage and Power Trade-off Across Model Sizes.} For base and small models, 32\,KB covers about \SI{66.5}{\percent} of kernels, whereas tiny reaches \SI{93.8}{\percent}. Larger LMMs reduce CPU fallback, but the added static power weakens the PDP gain under the current \SI{28}{\nm} synthesis assumptions.
\item \textbf{PDP Comparison under a TDP-Based Power Model.} From FPGA measurements and \SI{28}{\nm} ASIC projections, the co-designed IMAX configuration achieves a PDP \num{2.35}$\times$ lower than Jetson AGX Orin and \num{10.48}$\times$ lower than RTX~4090 for Whisper-tiny.en Q8\_0 under a TDP-based comparison.
\end{itemize}

The remainder of this paper is organized as follows. 
Section~\ref{rwork} surveys related work in AI accelerators and describes the IMAX3 architecture.
Section~\ref{proposed} details the implementation and co-design decisions.
Section~\ref{ex_and_re} presents the experimental results and analysis. 
Section~\ref{discussion} offers further design-factor analysis and discusses the main limitations of the present evaluation, and Section~\ref{conclusions} concludes the paper.

\newpage  
\section{Related Work}
\label{rwork}

Hardware acceleration for AI inference in this paper falls into three categories: task-dedicated ASR accelerators, general-purpose platforms such as FPGAs and edge GPUs, and reconfigurable array architectures. This section reviews representative designs in each category and places the IMAX-based implementation in that context.

\subsection{ASR-Specific Dedicated Hardware Accelerators}

Deep neural network-based ASR places high throughput and low-latency requirements on inference hardware, motivating task-specific accelerator designs. These accelerators achieve low energy per inference within a single model topology, but their fixed-function nature limits reuse as ASR model architectures change between generations.

Park et al.\nobreak\cite{park2022lowlatency} built a hybrid system that places a CNN on an FPGA and a language model on a smartphone SoC for low-latency keyword spotting. Because the datapath is bound to a fixed model topology, it cannot be repurposed for encoder-decoder architectures such as Whisper.
Yamini et al.\nobreak\cite{yamini2023hardware} proposed a systolic array accelerator for a Transformer-based end-to-end ASR system, exploiting the regularity of fixed-size matrix multiplications. Systolic arrays, however, lose utilization on the variable-length dimensions of Whisper's attention. The encoder sequence length depends on input audio duration, and autoregressive decoding produces queries of length one per step.

Other accelerator work targets non-Transformer ASR models. Yin et al.\nobreak\cite{spiking_lstm} implemented a spiking LSTM on an FPGA, reducing dynamic power through event-driven computation. Hu et al.\nobreak\cite{Hu_2022} designed a dedicated FPGA datapath for graph convolutional neural networks~(GCNNs), and Lu et al.\nobreak\cite{multi_task} proposed a multi-task accelerator for multiple fixed DNN topologies.
These designs share a common limitation. The hardware optimized for one model generation requires re-engineering when the target architecture changes.

This problem is acute in ASR. OpenAI has released multiple Whisper generations with different encoder layer counts, attention head counts, and hidden dimensions. A fixed-function accelerator targeting one Whisper variant would require full hardware re-engineering for the next. A programmable architecture accommodates such changes through software alone, without hardware redesign.
In contrast to these dedicated accelerators, our work evaluates a programmable architecture on an open encoder-decoder ASR model using a reusable datapath.

\subsection{FPGA and GPU-Based General Accelerators}

FPGAs and edge GPUs offer more flexibility than ASICs while delivering higher throughput per watt than CPUs.

On the FPGA side, Zhang et al.\nobreak\cite{fpgacnn} used roofline model analysis to match CNN accelerator throughput to available memory bandwidth. Li et al.\nobreak\cite{li2020ftransenergyefficientaccelerationtransformers} extended this direction to Transformers with FTrans, an FPGA-based accelerator that uses block-circulant weight representation to compress the model and reduce multiply-accumulate count. Lu et al.\nobreak\cite{lu2020hardwareacceleratormultiheadattention} designed pipelined hardware for multi-head attention and feed-forward layers.
Tambe et al.\nobreak\cite{EdgeBERT} developed EdgeBERT, a 12\,nm ASIC that integrates entropy-based early exit, adaptive attention span, and floating-point quantization, achieving up to 53$\times$ lower energy than a mobile GPU on BERT tasks. EdgeBERT targets a fixed encoder-only Transformer topology and does not transfer directly to encoder-decoder models such as Whisper, where the encoder processes fixed-length audio features and the autoregressive decoder generates variable-length token sequences.

LLM inference has driven further FPGA work. Chen et al.\nobreak\cite{llm4fpga} showed that decoupled spatial dataflow on FPGAs can outperform GPUs on memory-bound LLM workloads. Zeng et al.\nobreak\cite{zeng2024flightllm} introduced FlightLLM, an FPGA mapping flow that exploits weight sparsity and reconfigurable DSP chains. Xu et al.\nobreak\cite{llama2fpga,llamaf} proposed Llama2 accelerators for embedded FPGAs targeting IoT deployments. These FPGA solutions, however, rely on model-specific bitstream generation and fixed dataflow optimization, requiring re-engineering when the model architecture changes.

On the GPU side, edge platforms such as the NVIDIA Jetson AGX Orin\nobreak\cite{nvidia_jetson_agx_orin} draw 15--45\,W, which may exceed the budget of battery-operated or thermally constrained devices. GPU throughput also scales with batch size, whereas single-utterance ASR inference is inherently batch-size-one.
These platforms share our evaluation setting of single-utterance local execution. Our work differs from them in target metric. We study whether a programmable CGLA can reduce PDP for batch-size-one Whisper inference under the same local-execution condition.

\subsection{CGLA Architecture and IMAX}
\label{subsec:cgla_imax}

CGRAs map dataflow graphs onto a spatial PE array, combining programmability with lower energy per operation than GPUs\nobreak\cite{cgrasurvey}.
Conventional 2D CGRA meshes, however, incur compilation overhead because inter-PE routing across a 2D mesh becomes a bottleneck at scale\nobreak\cite{cgra1}. Domain-specific extensions, such as the lightweight-CNN CGRA of Lee et al.\nobreak\cite{cgra_cnn2}, reduce this overhead for particular workloads at the cost of generality. Wijerathne et al.\nobreak\cite{hmap} proposed hierarchical abstraction to accelerate CGRA placement and routing, but the trade-off between mapping flexibility and compilation cost remains.

IMAX\nobreak\cite{imax_access} avoids this problem by arranging PEs and LMMs in a one-dimensional linear array rather than a 2D mesh. The linear topology eliminates placement-and-routing search: the compiler maps C++ loop bodies directly onto the PE chain as a dataflow graph, and the resulting pipeline matches dot-product operations.

\begin{figure*}[t]
  \centering
  \includegraphics[width=1.9\columnwidth]{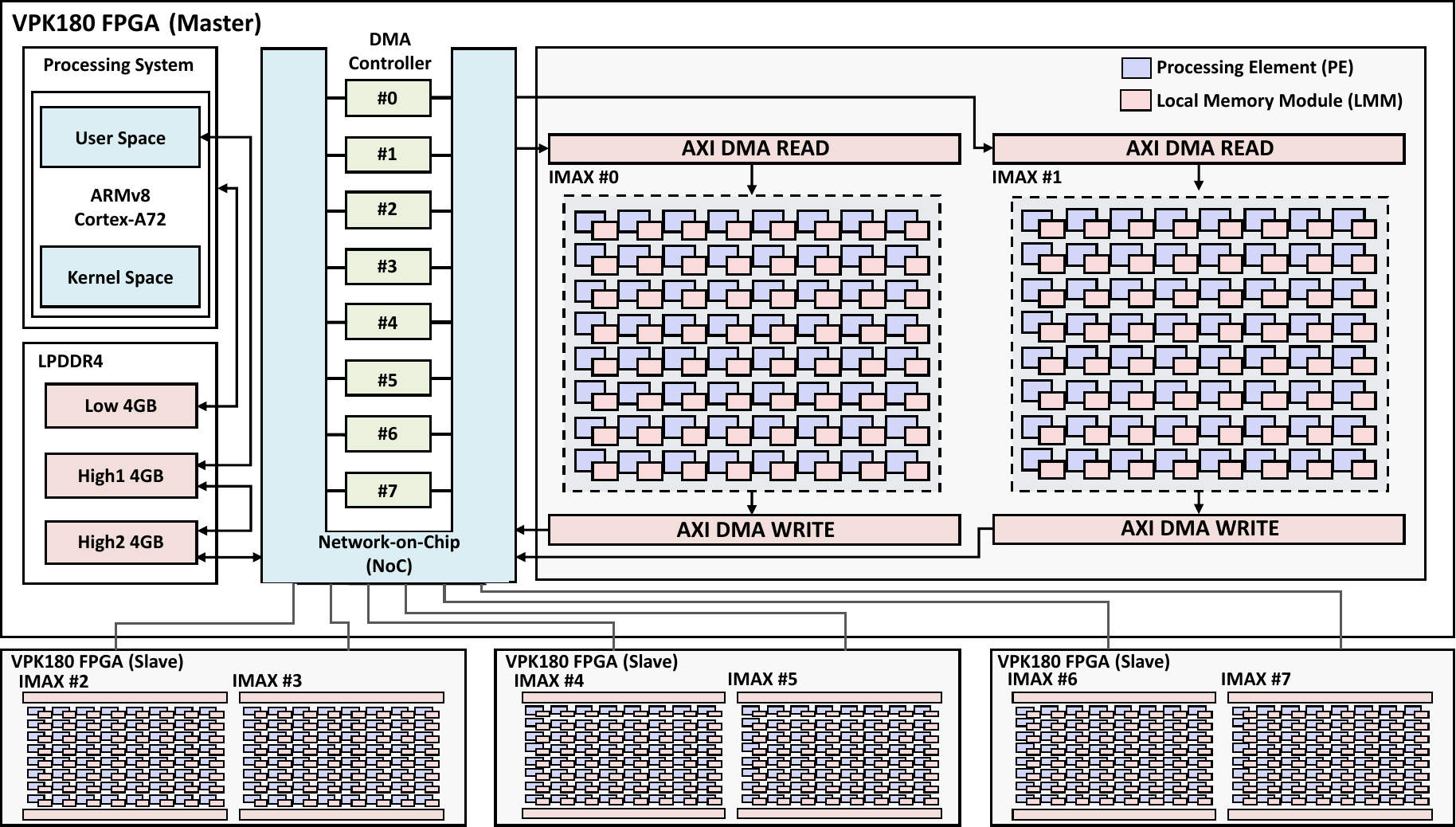}
  \caption{High-level overview of the IMAX3 multi-FPGA prototype, consisting of a host processor, a PCIe bridge board, and four AMD Versal VPK180 boards that implement the accelerator.}
  \label{fig:imax3_conf}
\end{figure*}

Fig.~\ref{fig:imax3_conf} provides a high-level overview of IMAX3, the third generation of the IMAX family, implemented on an AMD Versal VPK180 FPGA.
IMAX3 is organized as a heterogeneous SoC comprising a PS with a dual-core ARM Cortex-A72 processor running Linux, and a PL hosting the accelerator cores.
The PS is connected to an 8\,GB LPDDR4 memory subsystem, of which 4\,GB serves as the OS and application buffer and 4\,GB as the DMA buffer for host--accelerator data exchange.
A high-bandwidth NoC bridges the PS and PL domains, and a dedicated DMA controller provides eight independent channels, each mapped one-to-one to an IMAX compute lane, eliminating bus contention and allowing concurrent data loading across all active lanes.
Although the FPGA supports eight lanes, our current evaluation uses a two-lane configuration because the dual-core Cortex-A72 host becomes the scheduling bottleneck beyond two lanes\nobreak\cite{eto2025implementation}.
Scaling to the full eight-lane capacity requires upgrading the host to a higher-core-count processor, which is left as future work.

\begin{figure}[t]
  \centering
  \includegraphics[width=0.9\columnwidth]{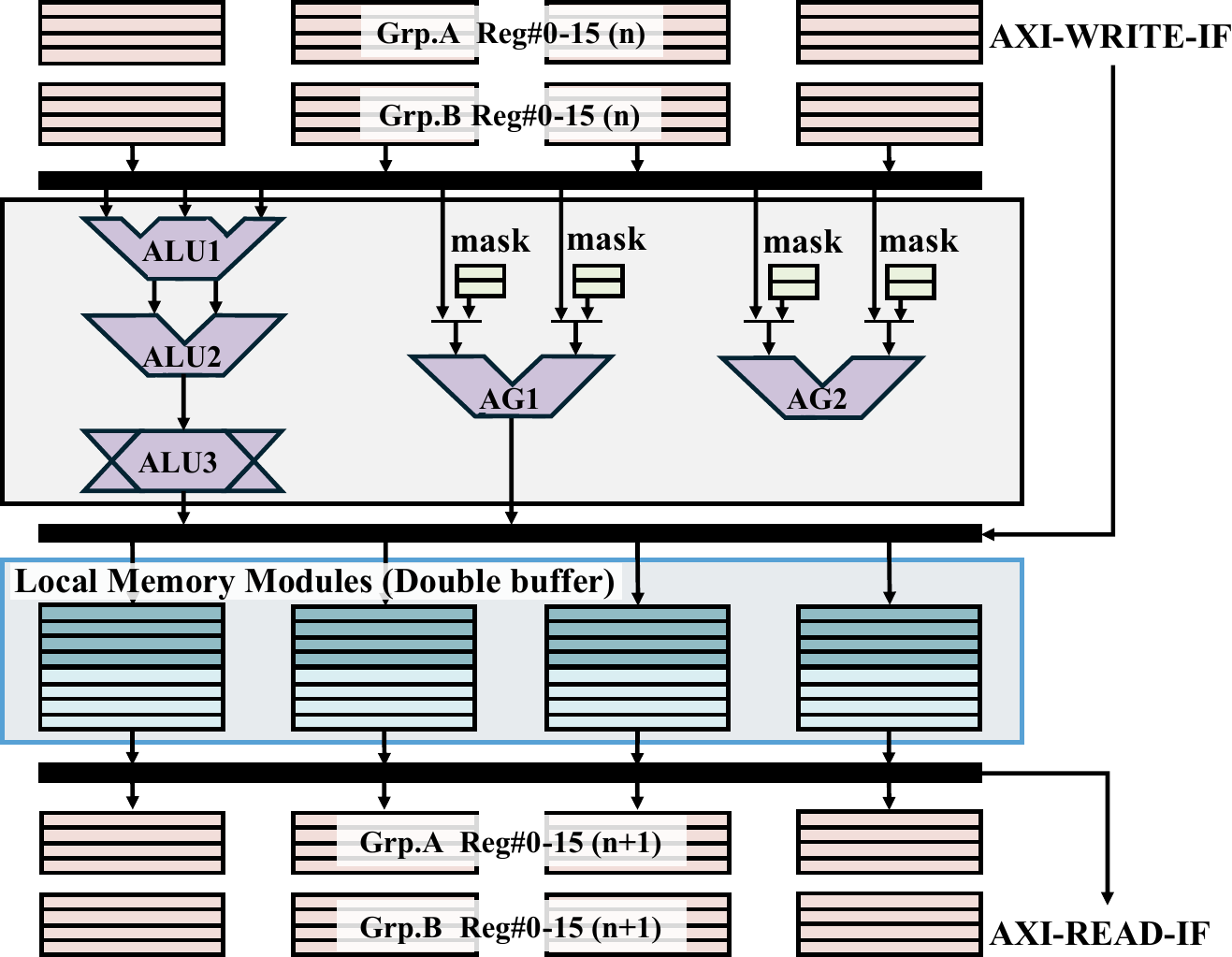}
  \caption{Internal structure of an IMAX compute lane.
    PEs and LMMs are arranged alternately in a one-dimensional linear array.
    The execution data path (blue) carries partial results between adjacent PEs,
    while the memory data path (red) connects each PE to its neighboring LMMs.}
  \label{fig:pe_detailed}
\end{figure}

The internal structure of each compute lane is depicted in Fig.~\ref{fig:pe_detailed}.
Each lane implements 64~PEs and their associated LMMs arranged alternately in a strict one-dimensional linear array.
Confining inter-PE communication to adjacent neighbors along a single axis produces deterministic scheduling and eliminates placement-and-routing search. The IMAX compiler translates C++ loop bodies directly onto the PE array as a dataflow graph.
Two data paths operate concurrently within each lane. One is an execution data path in which partial results pass directly from each PE to its downstream neighbor, forming a feed-forward pipeline. The other is a memory data path in which each PE directly accesses its adjacent LMMs with single-cycle latency.

Each PE is a multi-functional compute unit comprising three specialized ALUs, two AGs, a dual-group register file (16~registers per group), and a double-buffered LMM.
ALU1 handles integer arithmetic including multiply-accumulate operations; ALU2 performs bitwise operations and bit-field extraction, enabling in-line FP16-to-FP32 conversion without dedicated conversion hardware; and ALU3 provides barrel-shift and rotate capabilities for dequantization.
The two AGs compute memory addresses in parallel with ALU operations, preventing address calculation from consuming ALU cycles.
The LMM employs hardware-managed double-buffering. While one bank supplies operands to the PE, the DMA controller simultaneously prefetches the next data block into the other bank, completely overlapping DMA transfer with PE computation and masking memory latency.

Prior IMAX studies have mapped SpGEMM and FFT\nobreak\cite{imax_access}, CNN-based image segmentation\nobreak\cite{unetimax,imaxcnn3}, general CNN inference\nobreak\cite{imaxcnn2}, LLM inference\nobreak\cite{uetanicgra,eto2025implementation}, approximate k-NN search for RAG\nobreak\cite{kim-knn}, and lightweight cryptography\nobreak\cite{cgra_crypto} onto the same PE array, confirming that the CGLA datapath is not restricted to a single workload class.

This paper evaluates IMAX3 on ASR, targeting the Whisper family of encoder-decoder Transformer models.
Unlike the task-specific accelerators reviewed earlier in this section, our implementation can be adapted to different Whisper model variants through software modifications alone.
Two differences separate this work from the studies above. First, we use a programmable CGLA rather than a fixed-function accelerator tied to a single ASR topology. Second, we focus on the dominant Whisper kernel under controlled local execution to isolate the hardware-level trade-off between kernel coverage, latency, and energy.
We focus on the dot-product operation, which our profiling reveals to be the dominant computational kernel, accounting for over 87\% of total CPU execution time.
Section~\ref{proposed} details the kernel mapping and co-design decisions for Whisper.

\section{Implementation of whisper.cpp on IMAX}
\label{proposed}

This section details the hardware/software co-design for executing the principal Whisper kernels on IMAX3 (Section~\ref{subsec:cgla_imax}).
As illustrated in Fig.~\ref{fig:whisper_process}, the whisper.cpp \texttt{ggml\_mul\_mat()} calls (GEMM and dot-product), which account for over 87\,\% of total CPU execution time, are offloaded to IMAX via DMA, while all remaining operations (Softmax, Layer Normalization, etc.\@) remain on the host CPU.

\subsection{Implementation Substrate and Design Objectives}
The implementation substrate is whisper.cpp\nobreak\cite{ggerganov_whisper.cpp}.
Its minimal dependencies and flat memory layout avoid the overhead of high-level runtime layers, giving direct access to the hardware interface. Our evaluation targets the FP16 and Q8\_0 execution paths of the Whisper-tiny.en model. Q8\_0 represents model weights in 8-bit integers\nobreak\cite{dettmers2022llmint88bitmatrixmultiplication,frantar2023optq}, cutting the memory footprint by roughly 2$\times$ relative to FP16 and replacing floating-point multiply-accumulate with integer arithmetic. Q8\_0 and FP16 produce matching WER on standard English ASR benchmarks under whisper.cpp's default decoding settings\nobreak\cite{ggerganov_whisper.cpp}. No new quantization is proposed here; FP16 and Q8\_0 are used without modification from whisper.cpp, and the evaluation focuses on how each path maps onto the IMAX datapath.

Mapping Whisper onto IMAX requires co-design along two axes.
Our design has two goals.
First, we increase computational throughput by exploiting SIMD execution and column-wise multithreading.
Second, we increase local-memory coverage by stripping alignment padding and packing operands densely before DMA.

\begin{figure}[t]
  \centering
  \includegraphics[width=1.0\columnwidth]{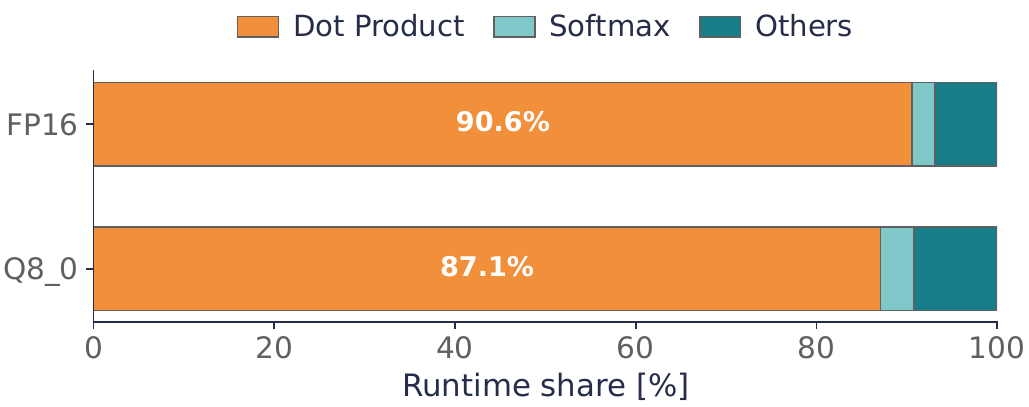}
  \caption{Runtime-share breakdown of Whisper-tiny.en inference on ARM Cortex-A72 (2 threads, audio duration $\approx$ 10\,s). The dot-product kernel dominates both FP16 and Q8\_0 execution.}
  \label{fig:runtime_breakdown}
\end{figure}

Fig.~\ref{fig:runtime_breakdown} summarizes the CPU execution-time breakdown for Whisper-tiny.en on the host ARM Cortex-A72, measured with two threads.
The dot-product operation dominates, accounting for \SI{90.6}{\percent} of total CPU execution time for the FP16 model and \SI{87.1}{\percent} for the Q8\_0 model.
All remaining operations, including Softmax, Layer Normalization, and others, collectively account for less than \SI{13}{\percent}.
The dot-product kernel is therefore the primary offload target. Offloading it to IMAX reduces the dominant CPU-side workload and lowers end-to-end latency and PDP.

Table~\ref{tab:whisper_arch} summarizes the principal architectural parameters of the Whisper-tiny.en model that are most relevant to the subsequent kernel mapping. The model dimension, feed-forward dimension, and per-head attention dimension are all multiples of 16, which later supports the burst-length choice discussed in Section~\ref{subsec:burst}.

\begin{table}[t]
    \centering
    \caption{Architectural parameters of the Whisper-tiny.en model. These dimensions determine the matrix sizes for the \texttt{ggml\_mul\_mat()} kernels offloaded to the IMAX accelerator.}
    \label{tab:whisper_arch}
    \begin{tabularx}{0.9\columnwidth}{>{\raggedright\arraybackslash}X r}
      \toprule
      \textbf{Parameter} & \textbf{Value} \\
      \midrule
      Transformer model dimension ($d_\text{model}$) & 384 \\
      Number of attention heads ($n_\text{heads}$) & 6   \\
      Per-head attention dimension ($d_k = d_\text{model}/n_\text{heads}$) & 64  \\
      Feed-forward hidden dimension ($d_\text{ff}$) & 1536 \\
      Number of encoder layers & 4   \\
      Number of decoder layers & 4   \\
      Input mel-spectrogram dimension & 80  \\
      Output vocabulary size & 51,864 \\
      \bottomrule
    \end{tabularx}
\end{table}

\subsection{Kernel-Level Co-design}
\label{subsec:burst}

We introduce a new FP16 dot-product kernel while reusing the Q8\_0 kernel from the previous work\nobreak\cite{eto2025implementation}.

\begin{figure}[t]
    \centering
    \includegraphics[width=0.9\columnwidth]{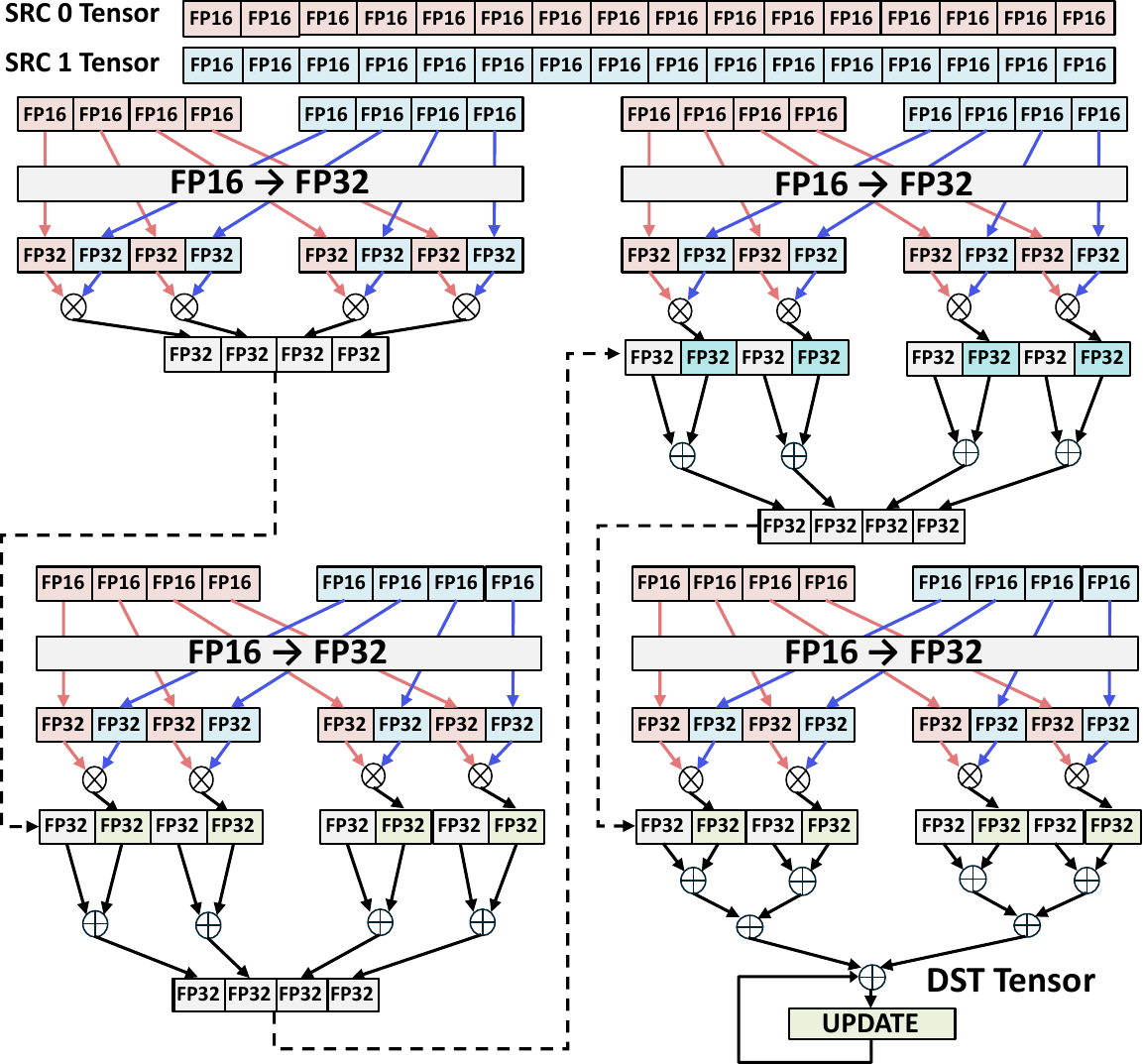}
    \caption{Processing flow of the FP16 kernel. FP16 operands are streamed from the LMMs, converted inline to FP32, accumulated with SIMD FMA operations, and reduced along the pipelined PE array without dedicated conversion hardware.}
    \label{fig:f16_kernel}
\end{figure}

The FP16 kernel exploits three features of the IMAX PE microarchitecture (Section~\ref{subsec:cgla_imax}).
First, ALU2's bit-manipulation unit converts FP16 to FP32 inline, avoiding dedicated conversion circuitry.
Second, we exploit two of IMAX's parallelization features to boost throughput.
\begin{itemize}
    \item \textbf{SIMD Operations.} Two 32-bit FMA operations execute concurrently on a single 64-bit ALU1 datapath, doubling the FMA throughput per PE.
    \item \textbf{Column-wise Multithreading.} Four logical FMA operations are time-multiplexed onto a single physical FPU, hiding the FPU pipeline latency.
\end{itemize}
The resulting FP16 processing flow is shown in Fig.~\ref{fig:f16_kernel}. Input operands are streamed from the LMMs, converted inline from FP16 to FP32, accumulated by SIMD FMA operations, and reduced along the lane through the pipelined PE array. The FP16 path reuses IMAX's native dataflow with no dedicated conversion hardware.

\begin{figure}[t]
  \centering
  \includegraphics[width=0.9\columnwidth]{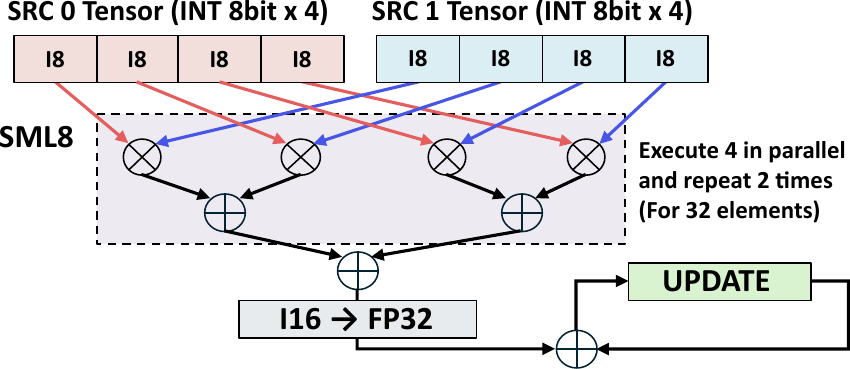}
  \caption{Processing flow of one unit of the Q8\_0 kernel. Quantized operands and scale data are streamed through the lane, processed by packed integer multiply-accumulate operations, and then reduced into the final accumulated output.}
  \label{fig:q8dot_one_unit}
\end{figure}

For the Q8\_0 path, we reuse the quantized dot-product kernel introduced in our previous work\nobreak\cite{eto2025implementation}. That earlier study provides the full implementation details, and we summarize only the key dataflow here because it is part of the present co-design. Fig.~\ref{fig:q8dot_one_unit} shows one representative processing unit of the Q8\_0 kernel. The kernel streams quantized operands and associated scaling data through the lane, performs parallel integer multiply-accumulate operations on packed data, and then reduces the partial sums before producing the final accumulated result. Reusing this Q8\_0 mapping lets the present study focus on the newly designed FP16 path while preserving the numerical behavior of the quantized baseline.

To realize these optimized processing flows, the following dedicated instructions have been added to the IMAX instruction set.
\begin{itemize}
\item \textbf{OP\_SML8.} A 2-way SIMD signed 8-bit integer multiply-add instruction. It independently multiplies the 8-bit segments of its input operands and sums the results into a sign-extended 32-bit output, enabling parallel processing of two INT8 multiply-accumulate operations on a single 64-bit datapath.
\item \textbf{OP\_AD32.} A 2-way 32-bit integer addition instruction used to aggregate the intermediate 32-bit results produced by OP\_SML8 along the pipeline.
\end{itemize}
Together, OP\_SML8 and OP\_AD32 let IMAX process GGML's packed quantized data structures without software unpacking loops. The Q8\_0 kernel occupies 46 of the 64 available PEs per lane, and the FP16 kernel uses 22. The remaining PEs stay idle per kernel invocation, but are available for other workloads, preserving programmability.

Whisper produces variable-length vectors that do not always align to the IMAX burst boundary.
IMAX processes fixed-length bursts, and handling residual elements would require conditional-store logic in the PE pipeline.
A mixed-execution split avoids this overhead.
Each vector is partitioned into a main segment whose length is a multiple of the burst length, offloaded to IMAX, and a short residual processed concurrently on a host CPU core.
IMAX never sees a partial burst, and the CPU handles only the tail elements.
Burst-length selection involves a three-way trade-off. A longer burst reduces per-invocation overhead but lowers the offload rate for short vectors and activates more PEs, raising power.
Burst lengths 8, 16, and 32 require 14, 22, and 38 active PEs per lane, respectively. Burst length 64 would exceed the 64-PE budget and was not implemented.
We select burst length 16 because it gives the best balance among alignment, PE cost, and residual work.
The principal computation dimensions in this model are the model dimension ($d_\text{model} = 384$), the feed-forward intermediate dimension ($d_\text{ff} = 1536$), and the per-head attention dimension ($d_k = 64$, derived as $d_\text{model} / n_\text{heads} = 384 / 6$).
All of these static dimensions are exact multiples of 16 (specifically, $384 = 24 \times 16$, $1536 = 96 \times 16$, and $64 = 4 \times 16$), ensuring that the main segments of operations on these dimensions are processed entirely by IMAX with zero residual elements on the CPU.
In the Whisper decoder, sequence length grows by one token per autoregressive step, so intermediate vector lengths are not necessarily multiples of 32 or 64.
Burst length 16 keeps the residual fraction small across the full range of encountered lengths.
Burst lengths 32 and 64 would reduce per-burst overhead on IMAX but increase the CPU-side residual for non-aligned lengths, worsening overall PDP.
From both alignment and hardware-cost viewpoints, burst length 16 yields the best trade-off among the evaluated options.
Section~\ref{subsec:burst_sensitivity} later validates this hypothesis experimentally by comparing burst lengths 8, 16, and 32 in terms of latency, PDP, and EDP.
In practice, each vector is decomposed into a main segment of length $\lfloor L/b \rfloor \times b$ and a short residual of length $L \bmod b$. The main segment is offloaded to IMAX, and the residual is handled concurrently on the host CPU. For Whisper-tiny.en, the dominant static GEMM dimensions ($d_k = 64$, $d_\text{model} = 384$, and $d_\text{ff} = 1536$) are all exact multiples of 16, so the CPU-side residual is zero for the principal kernels.

\subsection{Optimizing Data Handling and \\LMM Configuration}
Data transfer and LMM utilization form the second axis of the co-design.
FP16 tensors in whisper.cpp include padding for 32-byte alignment.
This padding wastes DMA bandwidth and occupies LMM capacity without contributing to computation.
The host CPU strips all padding and packs operands densely into the DMA buffer before offload.
Table~\ref{tab:padding_optimization_concise} shows the cumulative distribution of kernel memory footprints for the baseline (with padding) and our optimized implementation (without padding). 
As the table indicates, the baseline FP16 model can accommodate only \SI{1.39}{\percent} of kernels even with a 32\,KB LMM. 
After padding removal, a 32\,KB LMM covers \SI{93.80}{\percent} of all kernels, a 67$\times$ increase over the baseline.

\begin{table}[t]
    \centering
    \caption{Cumulative percentage of each model up to the specified LMM size limit}
    \label{tab:padding_optimization_concise}
    \setlength{\tabcolsep}{7pt}
    \begin{tabular}{l rr rr}
      \toprule
      \multirow{2}{*}{\textbf{LMM Limit}} & \multicolumn{2}{c}{\textbf{FP16 Model}} & \multicolumn{2}{c}{\textbf{Q8\_0 Model}} \\ \cmidrule(lr){2-3} \cmidrule(lr){4-5}
      & \textbf{Baseline} & \textbf{Optimized} & \textbf{Baseline} & \textbf{Optimized} \\
      \midrule
      \hspace{8pt}8KB   & 0.00\%   & 64.96\% & 0.00\%   & 64.96\% \\
      \hspace{4pt}16KB  & 1.39\%   & 66.35\% & 1.39\%   & 66.35\% \\
      \hspace{4pt}\textbf{32KB}  & 1.39\%   & \textbf{93.80\%} & 28.83\%  & \textbf{93.80\%} \\
      \hspace{4pt}64KB  & 93.81\%  & 93.80\% & 93.81\%  & 93.81\% \\
      128KB & 94.49\%  & 100.00\%& 97.24\%  & 100.00\% \\
      256KB & 100.00\% & 100.00\%& 100.00\% & 100.00\% \\
      \bottomrule
    \end{tabular}
\end{table}

\begin{figure}[t]
  \centering
  \includegraphics[width=1.0\columnwidth]{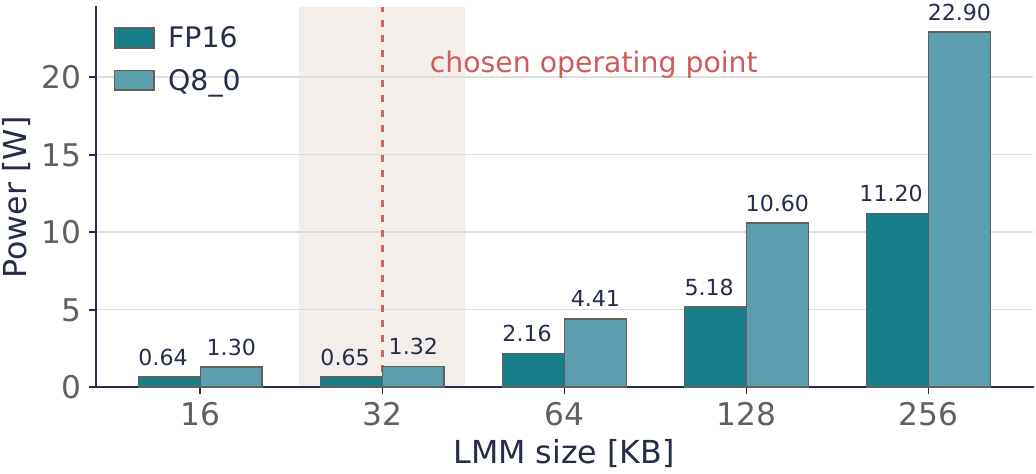}
  \caption{Power scaling by LMM size for FP16 and Q8\_0 kernels. The 32\,KB point is highlighted as the operating point used in the main evaluation.}
  \label{fig:lmm_size_power}
\end{figure}

\begin{table*}[t]
    \centering
    \begin{threeparttable}
    
    \caption{Physical specifications and performance comparisons of evaluated hardware platforms}
    \label{tab:processor_comparison_annotated}
    
    \begin{tabular}{@{} llrrrrrrr l @{}}
      \toprule
      \textbf{Device} & \textbf{CPU} & \textbf{Cores} & \textbf{Chip area} & \textbf{Process} & \textbf{Operating frequency} & \textbf{Memory} & \textbf{Power}\tnote{c} \\
      & & & \textbf{(mm$^2$)} & \textbf{node} & & & \textbf{(W)} \\
      \midrule
   
      \textbf{ARM Cortex-A72 (on Versal)} & - & 2 & - & 7\,nm & 1400\,MHz & 8\,GB DDR4 & 0.6485 \\
   
      \textbf{IMAX3 (Xilinx VPK180)} & ARM Cortex-A72 & 64\tnote{a} & - & 7\,nm & 140\,MHz\tnote{e} & 8\, + 4\,GB DDR4\tnote{b} & 180 \\
   
      \textbf{IMAX3 (28nm)} & - & 64\tnote{a} & 14.6 & \SI{28}{\nano\meter} & 840\,MHz& - &0.647 or 1.32 \\
   
      \textbf{Jetson AGX Orin 32GB} & Arm® Cortex-A78AE & 1792 & 200 & 8\,nm & 930\,MHz & 32\,GB LPDDR5 & 15\tnote{d} \\
   
      \textbf{NVIDIA RTX 4090} & Xeon W5-2455X & 16384 & 608 & 5\,nm & 2520\,MHz & 24\,GB DDR6 & 450 \\
      \bottomrule
    \end{tabular}

   \begin{tablenotes}[para,flushleft]
     \item[a] The number of cores for IMAX3 refers to the number of PEs per lane.
     \item[b] 8\,GB DDR4 for OS buffer and 4\,GB DDR4 for DMA buffer.
     \item[c] IMAX3~(28\,nm) is an estimated value, references for other devices are from Cortex-A72\nobreak\cite{Versal}, Jetson AGX Orin 32GB\nobreak\cite{nvidia_jetson_agx_orin}, NVIDIA RTX 4090\nobreak\cite{nvidia_ada}.
     \item[d] This device has multiple power consumption modes~(15--45\,W), and we executed in the lowest power consumption mode.
     \item[e] The IMAX FPGA operating frequency reported here corresponds to the measurement condition used for all experiments in this paper.
   \end{tablenotes}
    
     \end{threeparttable}
    \end{table*}  
Fig.~\ref{fig:lmm_size_power} presents per-LMM-size power estimates derived from logic synthesis using Synopsys Design Compiler with a TSMC \SI{28}{\nm} process library.
For the FP16 kernel, doubling the LMM from 16\,KB to 32\,KB raises power by only \SI{10}{\milli\watt} (from \SI{0.637}{\watt} to \SI{0.647}{\watt}).
Meanwhile, as shown in Table~\ref{tab:padding_optimization_concise}, expanding the LMM to 32\,KB substantially increases kernel coverage from \SI{66.35}{\percent} to \SI{93.80}{\percent}. 
The higher coverage reduces DRAM accesses and the associated dynamic power.
Based on this coverage-power analysis, 32\,KB was selected as the default LMM size for the main evaluation. 
Section~\ref{discussion} revisits this choice through a broader design-factor analysis that distinguishes latency reduction from the power cost of larger local memories.

\section{Experiments and Results}
\label{ex_and_re}
In this section, we evaluate latency and energy of the main whisper.cpp computation kernels on the IMAX3 accelerator.
Results from the FPGA prototype are reported alongside projected latency and power for a \SI{28}{\nm} ASIC derived from logic synthesis.

\subsection{Experimental Setup}
We evaluate three questions.
First, we compare IMAX against CPU and GPU platforms in end-to-end latency and PDP.
Second, we examine how that comparison changes from Whisper-tiny.en to Whisper-base.en and Whisper-small.en.
Third, we evaluate the burst-length trade-off within the IMAX design space.

The evaluation compares end-to-end~(E2E) latency and PDP across heterogeneous platforms for Whisper inference.
The latency and PDP workload for the detailed architectural sweep is whisper.cpp with its standard \texttt{jfk.wav} input (approximately 10\,s audio).
FP16 and Q8\_0 variants of Whisper-tiny.en are evaluated first under identical model and decoding conditions, because the operating-point proposal is derived from that workload.
The same evaluation flow then extends to Whisper-base.en and Whisper-small.en.
For the review-requested accuracy and multi-sample checks, we additionally evaluate 21 utterances from LibriSpeech test-clean\nobreak\cite{panayotov2015librispeech} using Whisper-tiny.en.
The utterances are LibriSpeech file IDs \texttt{6930-75918-0000} through \texttt{6930-75918-0020}.
This subset is used to check output agreement and per-sample latency, while the LMM-size and burst-length sweeps remain tied to the controlled \texttt{jfk.wav} workload.

IMAX was prototyped on an AMD Versal Premium VPK180 evaluation kit using Vivado 2024.1.
The SoC consists of a PS with a dual-core ARM Cortex-A72 CPU\nobreak\cite{Humrick_CortexA72_2016} and a PL hosting an 8-lane, 64-PE IMAX computation unit.
All FPGA measurements in this paper were conducted at \SI{140}{\mega\hertz} with one- and two-lane execution.
To evaluate future potential, we also report a \SI{28}{\nm} ASIC projection at \SI{840}{\mega\hertz}.
For comparison, we selected the embedded ARM Cortex-A72 CPU baseline, the NVIDIA Jetson AGX Orin edge GPU (lowest-power mode), and the NVIDIA GeForce RTX 4090 desktop GPU.
Table~\ref{tab:processor_comparison_annotated} is the primary hardware reference for the cross-platform comparison in this section and summarizes the evaluated platforms.
We standardized the software and operating system configurations to ensure a fair comparison.
\begin{itemize}
  \item The IMAX FPGA prototype runs a PetaLinux distribution, cross-compiled for the 64\,bit ARM (aarch64) architecture.
  \item The NVIDIA Jetson AGX Orin operates under the JetPack 6.1 software stack.
  \item The NVIDIA RTX 4090 system runs Ubuntu 22.04.5 LTS with CUDA Toolkit 12.4.
\end{itemize}
All platforms were benchmarked using an identical version of the whisper.cpp inference stack~\cite{ggerganov_whisper.cpp} and the exact same quantized model files, eliminating any variation due to software stack or model data.
The ASIC projection is derived from static timing and power analysis using Synopsys Design Compiler\cite{SynopsysNDDCUltra} with a TSMC \SI{28}{\nm} standard-cell library, including its standard cell libraries and wireload models, and assumed an average switching activity of 10\,\% for dynamic power estimation.
The FPGA prototype implements the full eight-lane fabric at 140\,MHz on a 7\,nm Versal device. The evaluation activates two lanes, yielding 128 active PEs.
The ASIC projection established a maximum achievable frequency of 840\,MHz, which is a 6$\times$ improvement over the FPGA prototype, with a single-lane chip area of 14.6\,mm$^2$~\cite{eto2025implementation}.
For the 32\,KB LMM configuration used in our main evaluation, the estimated one-lane power is 0.647\,W for FP16 kernels and 1.32\,W for Q8\_0 kernels, corresponding to 1.294\,W and 2.64\,W for two-lane operation, respectively.

To ensure fair software-level comparison, all platforms use the same inference stack, model files, and workload definition.
E2E latency was measured as wall-clock time on the host using \texttt{gettimeofday} with microsecond resolution.
For the \texttt{jfk.wav} cross-platform and design-space measurements, all reported metrics are the mean of 10 independent runs per setting with a fixed random seed.
The standard deviation remained below 3\,\% of the mean in those measurements.

For energy modeling, IMAX (28\,nm) power is estimated from synthesis and scaled by active-lane count, while GPU platforms are evaluated using nominal Thermal Design Power~(TDP) values.
We adopt TDP as the power metric for GPU platforms for two reasons.
First, TDP is a standardized, publicly available specification. A cross-architecture comparison on this basis requires no matched per-platform instrumentation.
Second, Whisper inference on GPUs is a compute-intensive workload that fully engages the GPU's arithmetic units during the GEMM-dominated phases. Under such conditions, actual power draw approaches TDP, making TDP a reasonable upper-bound estimate rather than an arbitrary worst case.
In GPU environments, CPU-bound phases are modeled as host CPU power (Xeon W-2455X)\nobreak\cite{Intel_Xeon_w5-2455X_Specs} plus GPU idle power.
TDP-based PDP values should be interpreted as architecture-level potential under peak-power assumptions. A comparison using directly measured time-averaged power (e.g., via wall-meter or on-board sensors) is left as future work.

\subsection{Accuracy and Multi-Utterance Output Check}

\begin{table}[t]
  \centering
  \caption{Transcript-level accuracy check on 21 LibriSpeech test-clean utterances using Whisper-tiny.en. WER and CER are measured against the LibriSpeech reference, and CPU--IMAX delta is measured between transcripts under the same precision path.}
  \label{tab:accuracy_check}
  \setlength{\tabcolsep}{4.5pt}
  \begin{tabular}{lccc}
    \toprule
    \textbf{Execution path} & \textbf{WER} & \textbf{CER} & \textbf{CPU--IMAX delta} \\
    \midrule
    CPU-only FP16 & 2.96\,\% & 0.74\,\% & -- \\
    IMAX FP16, 32\,KB LMM & 2.96\,\% & 0.74\,\% & 0.00\,\% \\
    CPU-only Q8\_0 & 5.03\,\% & 1.82\,\% & -- \\
    IMAX Q8\_0, 32\,KB LMM & 5.03\,\% & 1.82\,\% & 0.13\,\% \\
    \bottomrule
  \end{tabular}
\end{table}

Table~\ref{tab:accuracy_check} separates the output difference caused by the Q8\_0 model path from the output difference caused by IMAX offloading.
The IMAX FP16 row gives a 0.00\,\% mean normalized CPU--IMAX word-level transcript delta on this subset.
Compared with CPU-only FP16, CPU-only Q8\_0 changes the transcript by 3.40\,\% at the word level and 1.62\,\% at the character level on the same subset.
The IMAX Q8\_0 path matches the normalized CPU-only Q8\_0 transcript for 20 of the 21 utterances.
The remaining utterance differs by one normalized word, giving a mean CPU--IMAX word-level transcript delta of 0.13\,\% and a maximum per-utterance delta of 2.63\,\%.

The Q8\_0 scale and dequantization data are consumed in the same quantized model format as whisper.cpp.
We also quantify the scale/dequantization reconstruction error of that block format on the FP16 Whisper-tiny.en weight tensors.
Across 65 two-dimensional weight tensors and 36,430,848 scalar values, the reconstruction error is MAE $1.39{\times}10^{-4}$, RMSE $2.09{\times}10^{-4}$, maximum absolute error $3.41{\times}10^{-3}$, and relative L2 error $8.31{\times}10^{-3}$.
These checks report ASR output-level agreement and offline Q8\_0 reconstruction error.
They do not add a new quantizer or a new dequantization approximation to the IMAX path.

\begin{table}[t]
  \centering
  \caption{Multi-sample latency and transcript-agreement check on 21 LibriSpeech test-clean utterances using Whisper-tiny.en Q8\_0 with 32\,KB LMM. IMAX latency is the projected 28\,nm value including measured ARM fallback time.}
  \label{tab:multisample_latency_check}
  \scriptsize
  \begin{tabular*}{\columnwidth}{@{\extracolsep{\fill}}lccccc@{}}
    \toprule
    \textbf{ID} & \textbf{Dur.} & \textbf{CPU} & \textbf{IMAX} & \textbf{Speed} & \textbf{Delta} \\
    & \textbf{(s)} & \textbf{(s)} & \textbf{(s)} & & \textbf{(\%)} \\
    \midrule
    0000 & 3.5 & 9.99 & 9.34 & 1.07$\times$ & 0.00\,\% \\
    0001 & 14.2 & 12.83 & 12.74 & 1.01$\times$ & 0.00\,\% \\
    0002 & 5.0 & 10.08 & 9.42 & 1.07$\times$ & 0.00\,\% \\
    0003 & 23.3 & 14.90 & 15.26 & 0.98$\times$ & 0.00\,\% \\
    0004 & 11.1 & 11.88 & 11.40 & 1.04$\times$ & 0.00\,\% \\
    0005 & 13.2 & 12.01 & 11.76 & 1.02$\times$ & 0.00\,\% \\
    0006 & 5.8 & 10.63 & 9.94 & 1.07$\times$ & 0.00\,\% \\
    0007 & 3.3 & 9.95 & 9.23 & 1.08$\times$ & 0.00\,\% \\
    0008 & 4.8 & 10.20 & 9.62 & 1.06$\times$ & 0.00\,\% \\
    0009 & 7.3 & 11.23 & 10.89 & 1.03$\times$ & 0.00\,\% \\
    0010 & 3.0 & 9.71 & 9.11 & 1.07$\times$ & 0.00\,\% \\
    0011 & 3.2 & 10.09 & 9.41 & 1.07$\times$ & 0.00\,\% \\
    0012 & 1.9 & 9.68 & 8.99 & 1.08$\times$ & 0.00\,\% \\
    0013 & 2.9 & 9.86 & 9.26 & 1.06$\times$ & 0.00\,\% \\
    0014 & 16.8 & 13.23 & 13.20 & 1.00$\times$ & 0.00\,\% \\
    0015 & 6.4 & 10.78 & 10.29 & 1.05$\times$ & 0.00\,\% \\
    0016 & 10.0 & 11.70 & 11.37 & 1.03$\times$ & 0.00\,\% \\
    0017 & 6.2 & 10.67 & 10.17 & 1.05$\times$ & 0.00\,\% \\
    0018 & 10.8 & 11.58 & 11.27 & 1.03$\times$ & 0.00\,\% \\
    0019 & 11.6 & 12.01 & 11.68 & 1.03$\times$ & 0.00\,\% \\
    0020 & 14.4 & 13.01 & 12.85 & 1.01$\times$ & 2.63\,\% \\
    \midrule
    Mean & 8.5 & 11.24 & 10.82 & 1.04$\times$ & 0.13\,\% \\
    Std. & 5.5 & 1.37 & 1.63 & 0.03$\times$ & 0.56\,\% \\
    \bottomrule
  \end{tabular*}
\end{table}

Table~\ref{tab:multisample_latency_check} extends the single-audio setup with a public multi-utterance check.
Across the 21 utterances, the projected IMAX~(28\,nm) latency is close to the CPU-only Q8\_0 latency, with a mean speedup of 1.04$\times$ and one sample below parity.
The normalized CPU--IMAX transcript delta is 0.00\,\% for 20 utterances and 2.63\,\% for one utterance, giving a mean of 0.13\,\%.

\subsection{Cross-Platform Comparison Across Model Sizes}
Here, E2E latency is defined as the total wall-clock inference time for the complete transcription task.
PDP is the primary energy-efficiency metric, combining execution time and power consumption.
Unless otherwise noted, all cross-platform results in this subsection use the two-thread IMAX host configuration and the 32\,KB LMM setting.

\begin{figure*}[t]
  \centering
  \includegraphics[width=1.0\textwidth]{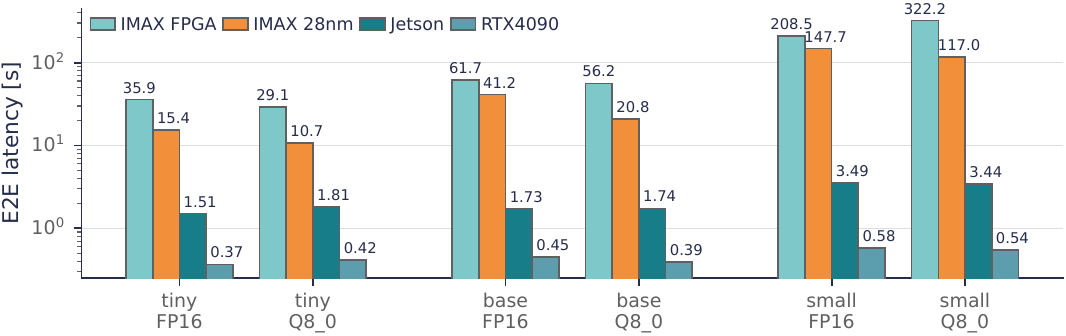}
  \caption{Cross-platform E2E latency comparison for Whisper tiny, base, and small models. IMAX FPGA values are measured on hardware at 140\,MHz. IMAX (28\,nm) values are projected from Synopsys Design Compiler synthesis at 840\,MHz and include no post-layout or silicon-variability derating. GPU latencies are wall-clock measurements under the software configurations described in Section~\ref{ex_and_re}.}
  \label{fig:cross_platform_latency}
\end{figure*}

\begin{figure*}[t]
  \centering
  \includegraphics[width=1.0\textwidth]{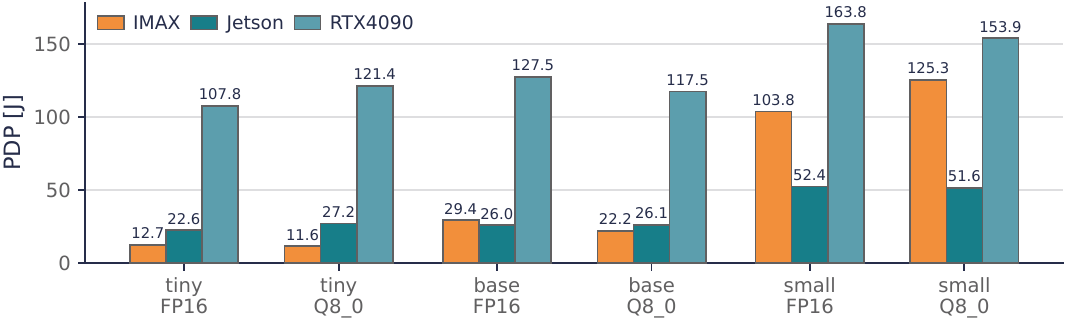}
  \caption{Cross-platform PDP comparison for Whisper tiny, base, and small models. PDP is computed as execution time multiplied by power. GPU power is nominal TDP, a standardized publicly available specification; IMAX (28\,nm) power is derived from Synopsys Design Compiler synthesis with TSMC standard cells at 10\,\% average switching activity. All PDP values are normalized peak-power estimates, not directly measured wall-plug energy. IMAX records the lowest PDP for Whisper-tiny.en under this model; the PDP gap over Jetson narrows as model size increases because the 32\,KB LMM covers fewer kernels for larger models.}
  \label{fig:cross_platform_pdp}
\end{figure*}

PDP is defined by Equation~\ref{eq:pdp}.
\begin{equation}
  \varPDP = \text{Execution time} \times \text{Power consumption}
  \label{eq:pdp}
\end{equation}

Lower PDP means less energy expended per completed inference.
For Jetson AGX Orin and RTX 4090, the power term uses nominal TDP, so the comparison reflects normalized peak-power conditions rather than measured workload-average power.
These comparisons are based on a TDP-normalized model and are intended as indicative rather than strictly equivalent cross-platform measurements.

Under the normalized power model used here, IMAX records the lowest PDP for Whisper-tiny.en, remains close to Jetson for Whisper-base.en Q8\_0, and no longer reduces PDP for Whisper-small.en.
As model size increases, the 32\,KB LMM leaves more kernels on the host CPU, and the PDP gap narrows.

Fig.~\ref{fig:cross_platform_latency} shows the raw latency gap across platforms.
The FPGA prototype is substantially slower than GPU platforms, as expected from its much lower clock frequency.
The projected IMAX~(28\,nm) ASIC narrows this gap considerably, especially for the tiny and base models, but still remains slower than Jetson and RTX~4090 in absolute E2E latency.
We separate raw latency from energy metrics before interpreting PDP.

Fig.~\ref{fig:cross_platform_pdp} summarizes the PDP trends for Whisper-tiny.en, Whisper-base.en, and Whisper-small.en under the normalized peak-power model described in Section~\ref{ex_and_re}, rather than under directly measured wall-plug energy conditions.
For Whisper-tiny.en, IMAX achieves its lowest PDP under this normalized comparison model.
The FP16 PDP of \SI{12.65}{\joule} is \num{1.79}$\times$ lower than Jetson's \SI{22.59}{\joule}, and the Q8\_0 PDP of \SI{11.58}{\joule} is \num{2.35}$\times$ lower than Jetson's \SI{27.16}{\joule}.
The same tiny-model Q8\_0 result is also \num{10.48}$\times$ lower than the RTX~4090 in PDP.
IMAX is slower than GPU platforms in absolute latency, but its PDP is 2.35$\times$ lower than Jetson's for Q8\_0, indicating a lower-energy and higher-latency trade-off.

\begin{figure*}[t]
  \centering
  \includegraphics[width=1.0\textwidth]{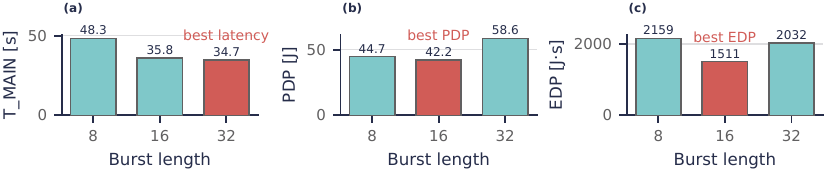}
  \caption{Burst-length trade-off for Whisper-tiny.en FP16 with 32\,KB LMM (2 lanes, 2 host threads). Burst length 32 minimizes latency, while burst length 16 yields the best PDP and EDP.}
  \label{fig:burst_tradeoff}
\end{figure*}

\begin{figure*}[t]
  \centering
  \includegraphics[width=1.0\textwidth]{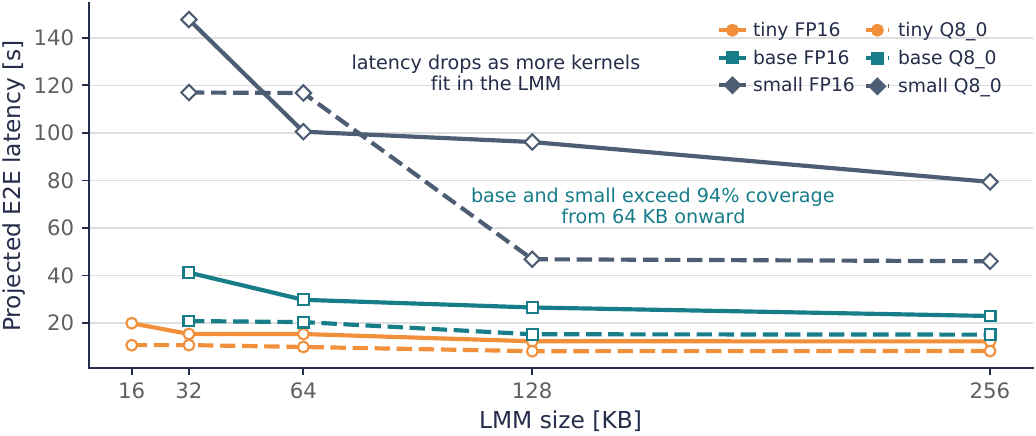}
  \caption{LMM-size effect on projected E2E latency across Whisper model sizes. Larger LMM capacities reduce CPU fallback by covering more kernels on IMAX. The 16\,KB points are shown for tiny.en only, because the new measurements were added to refine the tiny-model operating-point analysis.}
  \label{fig:lmm_tradeoff}
\end{figure*}
For Whisper-base.en, the PDP advantage narrows.
IMAX retains a modest PDP advantage for Q8\_0, with \SI{22.16}{\joule} versus Jetson's \SI{26.09}{\joule}, corresponding to a \num{1.18}$\times$ improvement.
For FP16, the two platforms are near parity, with IMAX at \SI{29.43}{\joule} and Jetson at \SI{25.98}{\joule}.
The 32\,KB LMM covers nearly all tiny-model kernels but falls short for base, where larger kernels increasingly trigger CPU fallback.
Simply enlarging local memory does not resolve this. The additional kernel coverage must be weighed against the static-power increase of the larger LMM (Section~\ref{subsec:lmm_power_scaling}).

For Whisper-small.en, the energy balance shifts in favor of Jetson.
The FP16 and Q8\_0 PDP values of IMAX rise to \SI{103.84}{\joule} and \SI{125.31}{\joule}, while Jetson records \SI{52.41}{\joule} and \SI{51.57}{\joule}, respectively.
The main cause is that the 32\,KB LMM no longer covers enough kernels, and enlarging the LMM under the current 28\,nm synthesis assumptions increases static power too aggressively.
Even in this regime, IMAX still remains substantially more favorable than the RTX~4090 in PDP across all evaluated models under the same normalized comparison model.

In summary, the present IMAX configuration records the lowest PDP for Whisper-tiny.en, remains close to Jetson for Whisper-base.en Q8\_0, and loses its PDP advantage for Whisper-small.en.
Under the current \SI{28}{\nm} synthesis assumptions, larger LMMs cover more kernels and reduce latency, but their static-power growth weakens the PDP gain.
Reducing the power increase of larger LMMs is the main design target for future work.

\subsection{Sensitivity Analysis of Burst Length}
\label{subsec:burst_sensitivity}

The burst-length sensitivity of the mixed-execution split is evaluated under the same Whisper-tiny.en FP16 setting with a 32\,KB LMM, two IMAX lanes, and two host CPU threads.
Burst lengths 8, 16, and 32 are compared.
Burst length 64 was not evaluated because the corresponding CGLA mapping would require more active PEs than are available in the current design.

Both PE count and power vary with burst length and must be accounted for alongside latency.
For one lane, the synthesized IMAX powers are 0.424\,W, 0.647\,W, and 1.09\,W for burst lengths 8, 16, and 32, corresponding to 14, 22, and 38 active PEs, respectively.
With two-lane execution, adding the ARM idle component yields system powers of 1.0967\,W, 1.5427\,W, and 2.4287\,W.
To compare these design points consistently, we derive the IMAX active time $\varT{active}$ from the sum of the measured \texttt{prompt\_eval} and \texttt{token\_gen} lane timings. $\varPDP_{\text{burst}}$ and $\varEDP_{\text{burst}}$ are defined by Equations~\ref{eq:pdp_burst} and~\ref{eq:edp_burst}.
\begin{equation}
  \varPDP_{\text{burst}} = \varT{active}\,\varP{IMAX} + (\varT{MAIN} - \varT{active})\,\varP{ARM},
  \label{eq:pdp_burst}
\end{equation}
\begin{equation}
  \varEDP_{\text{burst}} = \varPDP_{\text{burst}} \times \varT{MAIN}.
  \label{eq:edp_burst}
\end{equation}

Fig.~\ref{fig:burst_tradeoff} summarizes the burst-length trade-off.
Burst length 32 achieves the shortest wall-clock time, reducing $\varT{MAIN}$ to 34.7\,s, but this latency advantage comes at a much higher power cost because it activates 38 PEs per lane.
Its PDP and EDP rise to 58.6\,J and 2032.0\,J$\cdot$s, both well above the smaller-burst configurations.
At the other extreme, burst length 8 minimizes PE count, but the finer-grained execution increases $\varT{MAIN}$ to 48.3\,s, which raises the PDP and EDP to 44.7\,J and 2159.3\,J$\cdot$s, respectively.
Burst length 16 provides the best overall balance.
It achieves the lowest PDP at 42.2\,J and the lowest EDP at 1511.0\,J$\cdot$s, improving PDP by 5.6\% over burst length 8 and improving EDP by 30.0\% over burst length 8 and by 25.6\% over burst length 32.
Burst length 16 is the PDP- and EDP-optimal operating point among the three evaluated configurations.

\section{Discussion}
\label{discussion}
This section analyzes the main design factors (LMM sizing, burst-length selection, and computational utilization) and discusses the limitations of the present evaluation.

\subsection{Reducing Latency by Expanding LMM Coverage}

LMM size directly determines how much of the Whisper workload can remain on IMAX.
A small LMM causes frequent CPU fallback for kernels that do not fit in local memory, which directly increases end-to-end latency.
The analysis below first relates kernel coverage to projected latency, then explains why the latency-optimal LMM size does not coincide with the energy-optimal one.
Fig.~\ref{fig:lmm_tradeoff} plots latency versus LMM size. Larger LMM capacities reduce latency across all measured model sizes and both numerical formats.
The latency minimum occurs at 256\,KB for tiny FP16, base FP16, base Q8\_0, small FP16, and small Q8\_0, while tiny Q8\_0 reaches its minimum at 128\,KB.
The monotonic trend matches the kernel-coverage data in Table~\ref{tab:lmm_size_base_and_small}.

As more kernels fit inside the LMM, fewer operations fall back to the host CPU, and end-to-end latency decreases accordingly.
For base and small models, the 32\,KB LMM covers only about \SI{66.5}{\percent} of kernels (Table~\ref{tab:lmm_size_base_and_small}), whereas 64\,KB already covers more than \SI{94}{\percent}.
That increase in coverage corresponds to a clear latency reduction in Fig.~\ref{fig:lmm_tradeoff}.
From 32\,KB to 256\,KB, latency improves by \num{1.25}$\times$ for tiny FP16 and by \num{1.38}$\times$ for base Q8\_0.
For small Q8\_0, the latency drop is delayed until the LMM reaches 128\,KB.
The projected latency is 117.00\,s at 32\,KB, 116.81\,s at 64\,KB, 46.84\,s at 128\,KB, and 45.99\,s at 256\,KB.
The raw fallback counters explain why the 64\,KB point does not follow the aggregate coverage trend.
For small Q8\_0, the 64\,KB run still spends 35.8\,s in FP16 ARM fallback and 21.1\,s in Q8\_0 ARM fallback, while the 128\,KB run reduces these values to 9.0\,s and 0.0\,s, respectively.
Thus, although the 64\,KB LMM covers more kernels by count, the remaining fallback kernels are still expensive in time.
This makes the 64\,KB point insensitive to the aggregate coverage improvement and is consistent with FP16 being faster than Q8\_0 at the small-model 64\,KB point.
At 128\,KB, the Q8\_0 ARM fallback is eliminated in the measured counters, and Q8\_0 becomes faster.
Larger local memory covers more kernels on IMAX, and the reduced CPU fallback translates directly into lower latency.

  \begin{table}[t]
    \centering
    \caption{Cumulative dot-product kernel coverage (\%) for Whisper-tiny.en, base.en, and small.en as a function of LMM capacity. Bold values mark the first LMM size achieving over 93\% coverage for each model.}
    \label{tab:lmm_size_base_and_small}
    \setlength{\tabcolsep}{4.8pt}
    \begin{tabular}{lcccccc}
      \toprule
      \textbf{Model} & \textbf{16\,KB} & \textbf{32\,KB} & \textbf{64\,KB} & \textbf{128\,KB} & \textbf{256\,KB} \\
      \midrule
      tiny\hspace{5pt}(\,\,\,78\,MB) & 66.35\,\% & \textbf{93.80\,\%} & 93.80\,\% &100.00\,\% & 100.00\,\% \\
      base\hspace{3pt}(148\,MB) & 66.55\,\% & 66.54\,\% &\textbf{94.17\,\%} & 97.08\,\% & 99.89\,\% \\
      small\hspace{0.2pt}(488\,MB) & 66.53\,\% & 66.52\,\% &\textbf{94.36\,\%} & 96.89\,\% & 99.89\,\% \\
      \bottomrule
    \end{tabular}
  \end{table}

  \begin{figure}[t]
    \centering
    \includegraphics[width=1.0\columnwidth]{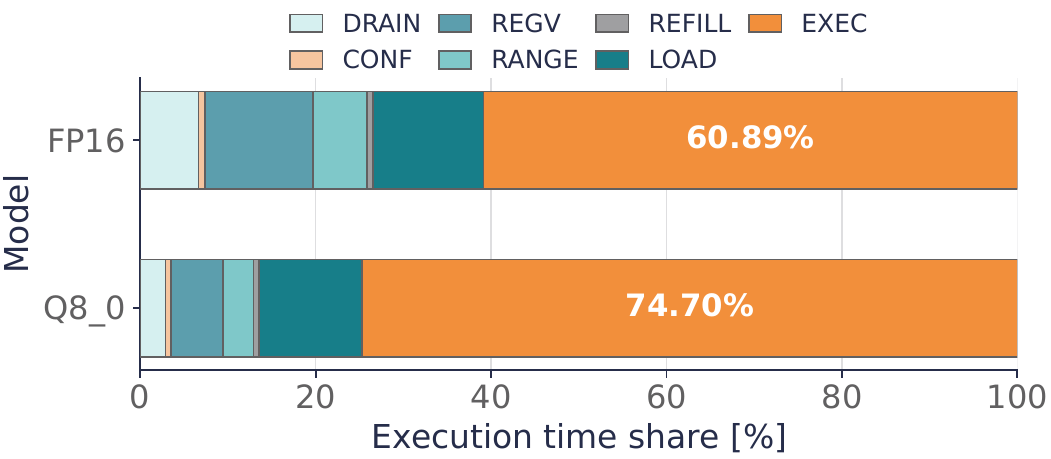}
    \caption{IMAX execution-time breakdown by kernel under the optimized configuration. EXEC occupies the majority of total time for both FP16 and Q8\_0, indicating that the offloaded kernel is predominantly compute-bound rather than limited by data movement or configuration overhead.}
    \label{fig:comparison_stacked_bar_chart}
  \end{figure}

\subsection{Implication of Burst-Length Selection}

The burst-length sensitivity study in Fig.~\ref{fig:burst_tradeoff} reveals a second design trade-off beyond LMM sizing.
Burst length 32 minimizes latency because its larger work granularity reduces execution overhead on IMAX.
That configuration, however, activates 38 PEs per lane, raising power enough to worsen both PDP and EDP.
Burst length 8 sits at the opposite extreme. It uses only 14 PEs per lane, but the finer-grained execution inflates runtime, again degrading PDP and EDP.
Burst length 16 balances these two effects.
It keeps the PE footprint moderate at 22 PEs per lane while preserving sufficiently coarse-grained execution, which is why burst length 16 minimizes both PDP and EDP among the three evaluated configurations.
Burst length must be chosen together with PE footprint and the workload's vector-length distribution. Minimizing latency alone or hardware cost alone yields a higher PDP.

\subsection{Execution-Time Breakdown and Compute Utilization}
We next examine how the co-designed kernels utilize the IMAX datapath.
Fig.~\ref{fig:comparison_stacked_bar_chart} presents a percentage breakdown of the kernel execution time components on IMAX under the optimized configuration. 
These components consist of EXEC (pure computation time on the PEs), LOAD/DRAIN (data transfer between DRAM and LMM), and CONF/REGV/RANGE/REFILL (IMAX configuration).
For the FP16 model, 60.89\% of the total execution time is spent in the EXEC component, while for the Q8\_0 model, this figure rises to 74.70\%. 
In contrast to many AI inference workloads that are memory-bound, these EXEC ratios indicate that IMAX operates in a compute-bound regime. The combination of padding removal, dense DMA packing, and double-buffered LMM access keeps data movement off the critical path.

With the offloaded kernel compute-bound, end-to-end latency is primarily bounded by the non-accelerated pipeline stages (less than 13\% of total CPU time), consistent with Amdahl's Law.

\subsection{Scalability to Larger Models}
\label{subsec:scalability_models}

The cross-platform evaluation in Fig.~\ref{fig:cross_platform_pdp} provides a supplementary view of how the tiny-oriented co-design behaves when applied to larger models.
Dot-product invocations grow from 477,153 (tiny) to 644,690 (base) and 1,920,955 (small).
The per-operation memory footprint, however, does not grow proportionally.
As shown in Table~\ref{tab:lmm_size_base_and_small}, at 32\,KB, only \SI{66.5}{\percent} of base and small kernels fit; at 64\,KB, over \SI{94}{\percent} fit.

Increasing kernel coverage directly reduces latency for larger models by shifting computation from the host CPU back to IMAX.
For all models tested, the practical turning point appears between 32\,KB and 64\,KB.
At 32\,KB, base and small models still leave about one third of kernels outside the LMM budget.
At 64\,KB, that uncovered fraction largely disappears for base and for the FP16 part of small, but the small Q8\_0 run still retains enough fallback time that its latency improvement appears only at 128\,KB.

For the larger models in this study, kernel coverage determines latency more directly than the execution model itself.
Extending the tiny-oriented design to larger models requires more local memory capacity together with higher host-side scheduling capacity.

\subsection{LMM size scaling and Architectural Implications}
\label{subsec:lmm_power_scaling}

Fig.~\ref{fig:lmm_size_power} plots power versus LMM size. Each doubling raises power while expanding the on-chip memory budget.
Combined with the coverage data in Table~\ref{tab:lmm_size_base_and_small}, this figure explains the main trade-off of the paper. Moving from 32\,KB to 64\,KB sharply reduces host-side fallback for Whisper-base.en and Whisper-small.en because kernel coverage rises from about \SI{66.5}{\percent} to more than \SI{94}{\percent}. That shift improves latency, but the larger LMM also raises static power.

For larger-model support, the design task is not simply to add more memory.
The design must add local-memory capacity without a disproportionate static-power penalty.
Under the current \SI{28}{\nm} synthesis, 32\,KB remains the PDP-optimal point for Whisper-tiny.en even though larger LMMs cut latency further.
In this design point, the PDP penalty comes from the power scaling of the current LMM cell rather than from a lack of latency benefit.

\subsection{Study Limitations}
\label{subsec:limitations}

Several limitations constrain the interpretation of these results. 
First, the IMAX~(28\,nm) vs.\ GPU energy comparisons use nominal TDP rather than directly measured time-averaged power. 
TDP is standardized, publicly available, and reproducible across architectures, but the resulting PDP values are normalized estimates, not exact workload-average energy. 
The reported Jetson and RTX comparisons reflect peak-power conditions, not wall-plug measurements under identical instrumentation. 
The 28\,nm ASIC figures likewise come from Synopsys Design Compiler synthesis with TSMC standard cell libraries, 10\,\% average switching activity, and no post-layout or silicon-variability derating. These numbers are directional estimates.

The experimental scope is also intentionally narrow. The detailed latency and PDP sweeps are based on a single 10-second audio clip (\texttt{jfk.wav}) with a fixed decoding configuration, which is sufficient for controlled comparison but does not capture variability across audio lengths, noise conditions, languages, or larger Whisper variants. 
The added LibriSpeech results in Tables~\ref{tab:accuracy_check} and~\ref{tab:multisample_latency_check} evaluate transcript agreement and per-sample latency across 21 public utterances, but they are not intended as a full ASR benchmark across speakers, acoustic conditions, or languages.
The accuracy check reports ASR output-level agreement and offline Q8\_0 reconstruction error.
In addition, we compare only platforms that can execute the same Whisper implementation under local conditions. Proprietary smartphone ASR stacks and cloud ASR services are excluded because they would mix model differences, closed software optimizations, and network or service-level effects with the hardware effects studied here. The results should therefore be read as a hardware-level comparison for open-model local inference rather than as a product benchmark against commercial ASR systems. For the same reason, the base and small results should be read as extensions of the tiny-model study, not as fully optimized deployment points in their own right.

Third, the current co-design favors energy over absolute throughput. 
Projected IMAX~(28\,nm) E2E latencies for Whisper-tiny.en are 15.39\,s (FP16) and 10.71\,s (Q8\_0) on an approximately 10\,s clip, so neither achieves real-time decoding. 
Only GEMM operations are offloaded, and non-linear operations remain on the host ARM Cortex-A72. 
As model size grows, the host-side share becomes more visible, and the dual-core CPU limits the evaluation to two of the eight available IMAX lanes. 
The paper is an architecture study of PDP scaling for the dominant kernel, not a claim of full real-time Whisper acceleration.

\section{Conclusion}
\label{conclusions}

We implemented and evaluated the principal computational kernels of the Whisper ASR model on the IMAX CGLA architecture.
We presented a co-design that combines an optimized FP16 dot-product kernel, an existing Q8\_0 kernel, SIMD execution, column-wise multithreading, and mixed execution for variable-length vectors.
For Whisper-tiny.en, 32\,KB LMM and burst length~16 form the operating point that jointly minimizes PDP and EDP under the current \SI{28}{\nm} power model.

The projected \SI{28}{\nm} IMAX ASIC achieves its lowest PDP on Whisper-tiny.en Q8\_0.
Under the normalized peak-power model used in this paper, that PDP is \num{2.35}$\times$ lower than Jetson AGX Orin and \num{10.48}$\times$ lower than RTX~4090.
For Whisper-base.en and Whisper-small.en, the PDP gap narrows and eventually reverses because 32\,KB local memory covers fewer kernels and larger LMMs raise static power under the current \SI{28}{\nm} synthesis assumptions.
These results indicate that the present IMAX design reduces PDP for Whisper-tiny.en under the evaluated conditions.
Future work will target larger-model support through lower-power LMM scaling and stronger host-side scheduling.

\bmsubsection*{Acknowledgments}
This work was supported by the JST-ALCA-Next Program (Grant Number JPMJAN23F4) and JSPS KAKENHI (Grant No. 22H00515). We also acknowledge the activities of VDEC, The University of Tokyo, in collaboration with NIHON SYNOPSYS G.K.

\bmsubsection*{Data Availability Statement}
The data that support the findings of this study are available from the corresponding author upon reasonable request.

\bmsubsection*{Conflicts of Interest}

The authors declare no conflicts of interest.

\bibliographystyle{wileyNJD-Chicago-lastoo}
\bibliography{wileyNJD-Chicago}

\begin{thebibliography}{47}
\providecommand{\natexlab}[1]{#1}
\providecommand{\url}[1]{\texttt{#1}}
\expandafter\ifx\csname urlstyle\endcsname\relax
  \providecommand{\doi}[1]{doi: #1}\else
  \providecommand{\doi}{doi: \begingroup \urlstyle{rm}\Url}\fi

\bibitem[\protect\citeauthoryear{Zhao, Huang, Xu, Lin, Liu, and Huang}{Zhao
  et~al.}{}]{expl_llm}
Andrew Zhao, Daniel Huang, Quentin Xu, Matthieu Lin, Yong-Jin Liu,, and Gao
  Huang. 2024.
\newblock ``{ExpeL: LLM agents are experiential learners}.''   In {\it
  Proceedings of the Thirty-Eighth AAAI Conference on Artificial Intelligence
  and Thirty-Sixth Conference on Innovative Applications of Artificial
  Intelligence and Fourteenth Symposium on Educational Advances in Artificial
  Intelligence}, AAAI'24/IAAI'24/EAAI'24, AAAI Press.

\bibitem[\protect\citeauthoryear{Goyal, Rastogi, Rajagopal, Yuan, Zhao,
  Chintagunta, Naik, and Ward}{Goyal et~al.}{}]{healai}
Sagar Goyal, Eti Rastogi, Sree~Prasanna Rajagopal, Dong Yuan, Fen Zhao, Jai
  Chintagunta, Gautam Naik,, and Jeff Ward. 2024.
\newblock ``{{HealAI}: A Healthcare LLM for Effective Medical Documentation}.''
    In {\it Proceedings of the 17th ACM International Conference on Web Search
  and Data Mining}, WSDM '24,   1167^^e2^^80^^931168.
\newblock New York, NY, USA: Association for Computing Machinery.

\bibitem[\protect\citeauthoryear{Li, Wang, Zeng, Wu, and Yang}{Li
  et~al.}{}]{llm_mas}
Xinyi Li, Sai Wang, Siqi Zeng, Yu~Wu,, and Yi~Yang.
\newblock
\newblock ``{{A survey on LLM-based multi-agent systems: workflow,
  infrastructure, and challenges}}.''  {\it Vicinagearth\/}1 (2024): 9.

\bibitem[\protect\citeauthoryear{Touvron, Lavril, Izacard, Martinet, Lachaux,
  Lacroix, Rozi^^c3^^a8re, Goyal, Hambro, Azhar, Rodriguez, Joulin, Grave, and
  Lample}{Touvron et~al.}{}]{touvron2023llamaopenefficientfoundation}
Hugo Touvron, Thibaut Lavril, Gautier Izacard, Xavier Martinet, Marie-Anne
  Lachaux, Timoth^^c3^^a9e Lacroix, Baptiste Rozi^^c3^^a8re, Naman Goyal, Eric
  Hambro, Faisal Azhar, Aurelien Rodriguez, Armand Joulin, Edouard Grave,, and
  Guillaume Lample. 2023.
\newblock ``LLaMA: Open and Efficient Foundation Language Models.''   ,
  arXiv:2302.13971.

\bibitem[\protect\citeauthoryear{Oruh, Viriri, and Adegun}{Oruh
  et~al.}{}]{ASR_1}
Jane Oruh, Serestina Viriri,, and Adekanmi Adegun.
\newblock
\newblock ``{Long Short-Term Memory Recurrent Neural Network for Automatic
  Speech Recognition}.''  {\it IEEE Access\/}10 (2022): 30069--30079.

\bibitem[\protect\citeauthoryear{Radford, Kim, Xu, Brockman, McLeavey, and
  Sutskever}{Radford et~al.}{}]{radford2022robustspeechrecognitionlargescale}
Alec Radford, Jong~Wook Kim, Tao Xu, Greg Brockman, Christine McLeavey,, and
  Ilya Sutskever. 2022.
\newblock ``{Robust Speech Recognition via Large-Scale Weak Supervision}.''   ,
  arXiv:2212.04356.

\bibitem[\protect\citeauthoryear{Abougarair}{Abougarair}{}]{ASR_smart_voice_assistant}
Ahmed Abougarair.
\newblock
\newblock ``Design and implementation of smart voice assistant and recognizing
  academic words.''  {\it International Journal of Robotics and Automation\/}8
  (2022): 27--32.

\bibitem[\protect\citeauthoryear{Wang, Han, Shafran, Wu, Chiu, Cao, Chen,
  Zhang, Soltau, Rubenstein, Zilka, Yu, Pundak, Siddhartha, Schalkwyk, and
  Wu}{Wang et~al.}{}]{ASR_T2T}
Mingqiu Wang, Wei Han, Izhak Shafran, Zelin Wu, Chung-Cheng Chiu, Yuan Cao,
  Nanxin Chen, Yu~Zhang, Hagen Soltau, Paul~K. Rubenstein, Lukas Zilka, Dian
  Yu, Golan Pundak, Nikhil Siddhartha, Johan Schalkwyk,, and Yonghui Wu. 2023.
\newblock ``{SLM: Bridge the Thin Gap Between Speech and Text Foundation
  Models}.''   In {\it 2023 IEEE Automatic Speech Recognition and Understanding
  Workshop (ASRU)},   1--8.

\bibitem[\protect\citeauthoryear{Luo, Zhou, Adelgais, and Zhang}{Luo
  et~al.}{}]{ASR_emergency_medicine}
Xiao Luo, Le~Zhou, Kathleen Adelgais,, and Zhan Zhang.
\newblock
\newblock ``{Assessing the Effectiveness of Automatic Speech Recognition
  Technology in Emergency Medicine Settings: a Comparative Study of Four
  AI-Powered Engines}.''  {\it Journal of Healthcare Informatics Research\/}~9,
  no. 3 (2025): 494--512.

\bibitem[\protect\citeauthoryear{Agency}{Agency}{}]{iea_energy_ai}
International~Energy Agency 2025.
\newblock ``{Energy and AI}.''
  \url{https://www.iea.org/reports/energy-and-ai}, Licence: CC BY 4.0,
  \url{https://creativecommons.org/licenses/by/4.0/}, Accessed on June 14,
  2025.

\bibitem[\protect\citeauthoryear{Shehabi, Smith, Hubbard, Newkirk, Lei, Siddik,
  et~al.}{Shehabi et~al.}{}]{shehabi2024united}
Arman Shehabi, Sarah Smith, Alan Hubbard, Adam Newkirk, Ning Lei, Md~Abu~Bakar
  Siddik, et~al. 2024.
\newblock ``2024 United States Data Center Energy Usage Report.''   Technical
  Report LBNL-2001637, Lawrence Berkeley National Laboratory.
\newblock Accessed on May 30, 2025.

\bibitem[\protect\citeauthoryear{Akabe, Trung Duong~LE, and Nakashima}{Akabe
  et~al.}{}]{imax_access}
Tomoya Akabe, Vu~Trung Duong~LE,, and Yasuhiko Nakashima.
\newblock
\newblock ``{IMAX}: A Power-Efficient Multilevel Pipelined CGLA and
  Applications.''  {\it IEEE Access\/}13 (2025): 31899--31911.

\bibitem[\protect\citeauthoryear{Han, Trimi, and Lee}{Han
  et~al.}{}]{HAN2026100674}
Hui Han, Silvana Trimi,, and Sang~M. Lee.
\newblock
\newblock ``Tiny Machine Learning (TinyML): Research trends and future
  application opportunities.''  {\it Array\/}29 (2026): 100674.

\bibitem[\protect\citeauthoryear{Park, Noh, Nam, Lee, and Park}{Park
  et~al.}{}]{park2022lowlatency}
Jaehyun Park, Hyeonkyu Noh, Hyunjoon Nam, Won-Cheol Lee,, and Hong-June Park.
\newblock
\newblock ``{A Low-Latency Streaming On-Device Automatic Speech Recognition
  System Using a CNN Acoustic Model on FPGA and a Language Model on
  Smartphone}.''  {\it Electronics\/}~11, no. 12.

\bibitem[\protect\citeauthoryear{Yamini, Mirishkar, Vuppala, and Purini}{Yamini
  et~al.}{}]{yamini2023hardware}
Shaarada~D. Yamini, Ganesh~S. Mirishkar, Anil~Kumar Vuppala,, and Suresh
  Purini. 2023.
\newblock ``{Hardware Accelerator for Transformer based End-to-End Automatic
  Speech Recognition System}.''   In {\it 2023 IEEE International Parallel and
  Distributed Processing Symposium Workshops (IPDPSW)},   93--100.

\bibitem[\protect\citeauthoryear{Yin, Dong, Chen, Ouyang, Wang, and Yang}{Yin
  et~al.}{}]{spiking_lstm}
Tingting Yin, Feihong Dong, Chao Chen, Chenghao Ouyang, Zheng Wang,, and
  Yongkui Yang.
\newblock
\newblock ``{A Spiking {LSTM} Accelerator for Automatic Speech Recognition
  Application Based on FPGA}.''  {\it Electronics\/}~13, no. 5.

\bibitem[\protect\citeauthoryear{Hu, Li, Wu, Li, and Chen}{Hu
  et~al.}{}]{Hu_2022}
Huaixiang Hu, Jiatong Li, Chunchun Wu, Xueyang Li,, and Yuping Chen.
\newblock
\newblock ``{Design and Implementation of Intelligent Speech Recognition System
  Based on FPGA}.''  {\it Journal of Physics: Conference Series\/}~2171, no. 1
  (2022): 012010.

\bibitem[\protect\citeauthoryear{Lu, Zhai, Saha, Ehsan, and McDonald-Maier}{Lu
  et~al.}{}]{multi_task}
Yufan Lu, Xiaojun Zhai, Sangeet Saha, Shoaib Ehsan,, and Klaus~D.
  McDonald-Maier. 2021.
\newblock ``{{FPGA} based Adaptive Hardware Acceleration for Multiple Deep
  Learning Tasks}.''   In {\it 2021 IEEE 14th International Symposium on
  Embedded Multicore/Many-core Systems-on-Chip (MCSoC)},   204--209.

\bibitem[\protect\citeauthoryear{Zhang, Li, Sun, Guan, Xiao, and Cong}{Zhang
  et~al.}{}]{fpgacnn}
Chen Zhang, Peng Li, Guangyu Sun, Yijin Guan, Bingjun Xiao,, and Jason Cong.
  2015.
\newblock ``Optimizing FPGA-based Accelerator Design for Deep Convolutional
  Neural Networks.''   In {\it Proceedings of the 2015 ACM/SIGDA International
  Symposium on Field-Programmable Gate Arrays}, FPGA '15,   161--170.
\newblock New York, NY, USA: Association for Computing Machinery.

\bibitem[\protect\citeauthoryear{Li, Pandey, Fang, Lyu, Li, Chen, Xie, Wan,
  Liu, and Ding}{Li
  et~al.}{}]{li2020ftransenergyefficientaccelerationtransformers}
Bingbing Li, Santosh Pandey, Haowen Fang, Yanjun Lyu, Ji~Li, Jieyang Chen, Mimi
  Xie, Lipeng Wan, Hang Liu,, and Caiwen Ding. 2020.
\newblock ``{FTRANS}: Energy-Efficient Acceleration of Transformers using
  FPGA.''   , arXiv:2007.08563.

\bibitem[\protect\citeauthoryear{Lu, Wang, Liang, Lin, and Wang}{Lu
  et~al.}{}]{lu2020hardwareacceleratormultiheadattention}
Siyuan Lu, Meiqi Wang, Shuang Liang, Jun Lin,, and Zhongfeng Wang. 2020.
\newblock ``{Hardware Accelerator for Multi-Head Attention and Position-Wise
  Feed-Forward in the Transformer}.''   , arXiv:2009.08605.

\bibitem[\protect\citeauthoryear{Tambe, Hooper, Pentecost, Jia, Yang, Donato,
  Sanh, Whatmough, Rush, Brooks, and Wei}{Tambe et~al.}{}]{EdgeBERT}
Thierry Tambe, Coleman Hooper, Lillian Pentecost, Tianyu Jia, En-Yu Yang, Marco
  Donato, Victor Sanh, Paul Whatmough, Alexander~M. Rush, David Brooks,, and
  Gu-Yeon Wei. 2021.
\newblock ``EdgeBERT: Sentence-Level Energy Optimizations for Latency-Aware
  Multi-Task NLP Inference.''   In {\it MICRO-54: 54th Annual IEEE/ACM
  International Symposium on Microarchitecture}, MICRO '21,
  830^^e2^^80^^93844.
\newblock New York, NY, USA: Association for Computing Machinery.

\bibitem[\protect\citeauthoryear{Chen, Zhang, Du, Xiang, Yue, Zhang, Cai, and
  Zhang}{Chen et~al.}{}]{llm4fpga}
Hongzheng Chen, Jiahao Zhang, Yixiao Du, Shaojie Xiang, Zichao Yue, Niansong
  Zhang, Yaohui Cai,, and Zhiru Zhang.
\newblock
\newblock ``{Understanding the Potential of FPGA-based Spatial Acceleration for
  Large Language Model Inference}.''  {\it ACM Trans. Reconfigurable Technol.
  Syst.\/}~18, no. 1.

\bibitem[\protect\citeauthoryear{Zeng, Liu, Dai, Yang, Fu, Wang, Ma, Sun, Li,
  Huang, Dai, Li, Wang, Zhang, Wen, Ning, and Wang}{Zeng
  et~al.}{}]{zeng2024flightllm}
Shulin Zeng, Jun Liu, Guohao Dai, Xinhao Yang, Tianyu Fu, Hongyi Wang, Wenheng
  Ma, Hanbo Sun, Shiyao Li, Zixiao Huang, Yadong Dai, Jintao Li, Zehao Wang,
  Ruoyu Zhang, Kairui Wen, Xuefei Ning,, and Yu~Wang. 2024.
\newblock ``{FlightLLM}: Efficient Large Language Model Inference with a
  Complete Mapping Flow on FPGAs.''   In {\it Proceedings of the 2024 ACM/SIGDA
  International Symposium on Field Programmable Gate Arrays}, FPGA '24,
  223--234.
\newblock New York, NY, USA: Association for Computing Machinery.

\bibitem[\protect\citeauthoryear{Xu, Wang, and Ji}{Xu et~al.}{}]{llama2fpga}
Han Xu, Xingyuan Wang,, and Shihao Ji. 2024.
\newblock ``{Towards Energy-Efficient Llama2 Architecture on Embedded FPGAs}.''
    In {\it Proceedings of the 33rd ACM International Conference on Information
  and Knowledge Management}, CIKM '24,   5570--5571.
\newblock New York, NY, USA: Association for Computing Machinery.

\bibitem[\protect\citeauthoryear{Xu, Li, and Ji}{Xu et~al.}{}]{llamaf}
Han Xu, Yutong Li,, and Shihao Ji. 2024.
\newblock ``{LlamaF}: An Efficient Llama2 Architecture Accelerator on Embedded
  FPGAs.''   In {\it 2024 IEEE 10th World Forum on Internet of Things
  (WF-IoT)},   1--7.

\bibitem[\protect\citeauthoryear{Karumbunathan}{Karumbunathan}{}]{nvidia_jetson_agx_orin}
Leela~S Karumbunathan 2022.
\newblock ``{Nvidia Jetson AGX Orin series}.''   , A Giant Leap Forward for
  Robotics and Edge AI Applications. Technical Brief.

\bibitem[\protect\citeauthoryear{Podobas, Sano, and Matsuoka}{Podobas
  et~al.}{}]{cgrasurvey}
Artur Podobas, Kentaro Sano,, and Satoshi Matsuoka.
\newblock
\newblock ``A Survey on Coarse-Grained Reconfigurable Architectures From a
  Performance Perspective.''  {\it IEEE Access\/}8 (2020): 146719--146743.

\bibitem[\protect\citeauthoryear{Torng, Pan, Ou, Tan, and Batten}{Torng
  et~al.}{}]{cgra1}
Christopher Torng, Peitian Pan, Yanghui Ou, Cheng Tan,, and Christopher Batten.
  2021.
\newblock ``{Ultra-Elastic CGRAs for Irregular Loop Specialization}.''   In
  {\it 2021 IEEE International Symposium on High-Performance Computer
  Architecture (HPCA)},   412--425.

\bibitem[\protect\citeauthoryear{Lee and Lee}{Lee and Lee}{}]{cgra_cnn2}
Jungi Lee, and Jongeun Lee.
\newblock
\newblock ``{Specializing CGRAs for Light-Weight Convolutional Neural
  Networks}.''  {\it IEEE Transactions on Computer-Aided Design of Integrated
  Circuits and Systems\/}~41, no. 10 (2022): 3387--3399.

\bibitem[\protect\citeauthoryear{Wijerathne, Li, Pathania, Mitra, and
  Thiele}{Wijerathne et~al.}{}]{hmap}
Dhananjaya Wijerathne, Zhaoying Li, Anuj Pathania, Tulika Mitra,, and Lothar
  Thiele.
\newblock
\newblock ``{HiMap}: Fast and Scalable High-Quality Mapping on CGRA via
  Hierarchical Abstraction.''  {\it IEEE Transactions on Computer-Aided Design
  of Integrated Circuits and Systems\/}~41, no. 10 (2022): 3290--3303.

\bibitem[\protect\citeauthoryear{Eto and Nakashima}{Eto and
  Nakashima}{}]{eto2025implementation}
Yu~Eto, and Yasuhiko Nakashima. 2026.
\newblock ``Implementation and Performance Analysis of LLaMA on a CGLA.''   In
  {\it Proceedings of the Fifth International Conference on Intelligent Systems
  and Networks},   516--525.
\newblock Singapore: Springer Nature Singapore.

\bibitem[\protect\citeauthoryear{Thi~Sang, Imamura, Akabe, and
  Nakashima}{Thi~Sang et~al.}{}]{unetimax}
Duong Thi~Sang, Ren Imamura, Tomoya Akabe,, and Yasuhiko Nakashima.
\newblock
\newblock ``{Energy Consumption Optimization of Multi-Dimensional U-Nets on
  CGLA}.''  {\it IEEE Access\/}13 (2025): 29476--29492.

\bibitem[\protect\citeauthoryear{Tanomoto, Takamaeda-Yamazaki, Yao, and
  Nakashima}{Tanomoto et~al.}{}]{imaxcnn3}
Masakazu Tanomoto, Shinya Takamaeda-Yamazaki, Jun Yao,, and Yasuhiko Nakashima.
  2015.
\newblock ``{A CGRA-Based Approach for Accelerating Convolutional Neural
  Networks}.''   In {\it 2015 IEEE 9th International Symposium on Embedded
  Multicore/Many-core Systems-on-Chip},   73--80.

\bibitem[\protect\citeauthoryear{Imamura, Guangxian, Thi, Pham, Zhang, and
  Nakashima}{Imamura et~al.}{}]{imaxcnn2}
Ren Imamura, Zhu Guangxian, Sang~Duong Thi, Hoai~Luan Pham, Renyuan Zhang,, and
  Yasuhiko Nakashima. 2024.
\newblock ``{Energy-Efficient 3D Convolution Using Interposed Memory
  Accelerator eXtension 2 for Medical Image Processing}.''   In {\it
  Proceedings of 2023 International Conference on Medical Imaging and
  Computer-Aided Diagnosis (MICAD 2023)}, edited by Ruidan Su, Yu-Dong Zhang,
  and Alejandro~F. Frangi,   62--71.
\newblock Singapore: Springer Nature Singapore.

\bibitem[\protect\citeauthoryear{Uetani and Nakashima}{Uetani and
  Nakashima}{}]{uetanicgra}
Hitoaki Uetani, and Yasuhiko Nakashima. 2024.
\newblock ``{ Implementation and Evaluation of LLM on a CGLA }.''   In {\it
  2024 Twelfth International Symposium on Computing and Networking (CANDAR)},
  252--258.
\newblock IEEE Computer Society.

\bibitem[\protect\citeauthoryear{Kim and Nakashima}{Kim and
  Nakashima}{}]{kim-knn}
Dohyun Kim, and Yasuhiko Nakashima.
\newblock
\newblock ``{Optimizing Matrix-Vector Operations With CGLA for High-Performance
  Approximate k-NN Search}.''  {\it IEEE Access\/}13 (2025): 111087--111097.

\bibitem[\protect\citeauthoryear{Thi, Luan~Pham, Duong~Le, Tran, Imamura,
  Nam~Nguyen, Tran, and Nakashima}{Thi et~al.}{}]{cgra_crypto}
Sang~Duong Thi, Hoai Luan~Pham, Vu~Trung Duong~Le, Thi~Diem Tran, Ren Imamura,
  Quoc~Duy Nam~Nguyen, Thi~Hong Tran,, and Yasuhiko Nakashima. 2023.
\newblock ``{Universal 32/64-bit CGRA for Lightweight Cryptography in Securing
  IoT Data Transmission}.''   In {\it 2023 IEEE 16th International Symposium on
  Embedded Multicore/Many-core Systems-on-Chip (MCSoC)},   419--425.

\bibitem[\protect\citeauthoryear{Gerganov}{Gerganov}{}]{ggerganov_whisper.cpp}
Georgi Gerganov 2023.
\newblock ``{whisper.cpp: Port of OpenAI's Whisper model in C/C++}.''
  \url{https://github.com/ggerganov/whisper.cpp}, Accessed: 2025-07-24.

\bibitem[\protect\citeauthoryear{Dettmers, Lewis, Belkada, and
  Zettlemoyer}{Dettmers et~al.}{}]{dettmers2022llmint88bitmatrixmultiplication}
Tim Dettmers, Mike Lewis, Younes Belkada,, and Luke Zettlemoyer. 2022.
\newblock ``LLM.int8(): 8-bit Matrix Multiplication for Transformers at
  Scale.''   , .

\bibitem[\protect\citeauthoryear{Frantar, Ashkboos, Hoefler, and
  Alistarh}{Frantar et~al.}{}]{frantar2023optq}
Elias Frantar, Saleh Ashkboos, Torsten Hoefler,, and Dan Alistarh. 2023.
\newblock ``{OPTQ}: Accurate Quantization for Generative Pre-trained
  Transformers.''   In {\it The Eleventh International Conference on Learning
  Representations}, .

\bibitem[\protect\citeauthoryear{Inc.}{Inc.}{}]{Versal}
Xilinx Inc. 2025.
\newblock ``Versal Power Demo.''
  \url{https://github.com/Xilinx/pm_demo/tree/master?tab=readme-ov-file},
  Accessed on May 30, 2025.

\bibitem[\protect\citeauthoryear{Corporation}{Corporation}{}]{nvidia_ada}
Nvidia Corporation 2023.
\newblock ``{NVIDIA ADA GPU ARCHITECTURE: Designed to deliver outstanding
  gaming and creating, professional graphics, AI, and compute performance}.''
  , Whitepaper.

\bibitem[\protect\citeauthoryear{Panayotov, Chen, Povey, and
  Khudanpur}{Panayotov et~al.}{}]{panayotov2015librispeech}
Vassil Panayotov, Guoguo Chen, Daniel Povey,, and Sanjeev Khudanpur. 2015.
\newblock ``LibriSpeech: An ASR corpus based on public domain audio books.''
  In {\it 2015 IEEE International Conference on Acoustics, Speech and Signal
  Processing},   5206--5210.

\bibitem[\protect\citeauthoryear{Humrick}{Humrick}{}]{Humrick_CortexA72_2016}
Mike Humrick 2016, January).
\newblock ``{ARM Cortex-A72 Architecture Deep Dive}.''
  \url{https://www.tomshardware.com/reviews/arm-cortex-a72-architecture,4424.html},
  Accessed on May 30, 2025.

\bibitem[\protect\citeauthoryear{Synopsys}{Synopsys}{}]{SynopsysNDDCUltra}
Inc. Synopsys 2024.
\newblock ``{DC Ultra: Concurrent Timing, Area, Power and Test Optimization}.''
  \url{https://www.synopsys.com/implementation-and-signoff/rtl-synthesis-test/dc-ultra.html},
  Accessed on May 25, 2025.

\bibitem[\protect\citeauthoryear{Corporation}{Corporation}{}]{Intel_Xeon_w5-2455X_Specs}
Intel Corporation 2025.
\newblock ``{Intel{\textregistered} Xeon{\textregistered} w5-2455X Processor
  (30M Cache, 3.20 GHz) Specifications}.''
  \url{https://www.intel.co.jp/content/www/jp/ja/products/sku/233420/intel-xeon-w52455x-processor-30m-cache-3-20-ghz/specifications.html},
  Accessed on May 30, 2025.

\end{thebibliography}

\end{document}